\documentclass[aps,prx,superscriptaddress,amsfonts,amsmath,amssymb,showpacs,floatfix,reprint,longbibliography]{revtex4-1}
\usepackage{url}
\usepackage{bm}
\usepackage{graphicx}
\usepackage{amsmath}
\usepackage{amstext}
\usepackage{amssymb}
\usepackage{amsfonts}
\usepackage{amsbsy}
\usepackage{verbatim}
\usepackage{color}
\usepackage[colorlinks=true, urlcolor=blue, linkcolor=blue, citecolor=blue, pdftex]{hyperref}
\usepackage{multirow}
\usepackage{float}
\usepackage{gensymb}
\usepackage{textcomp}
\usepackage{enumitem}
\usepackage[english]{babel}
\usepackage[pdftex]{epsfig}
\usepackage{ragged2e}
\usepackage{times}
\usepackage[table,xcdraw]{xcolor}
\usepackage{epstopdf}
\usepackage{physics}
\usepackage[export]{adjustbox}
\usepackage{dsfont}

\begin{document}
\title{Projective symmetry group classifications of quantum spin liquids on the simple cubic, body centered cubic, and face centered cubic lattices}
\author{Jonas Sonnenschein}
\email{j0nas@zedat.fu-berlin.de}
\affiliation{Dahlem Center for Complex Quantum Systems and Institut f\"ur Theoretische Physik, Freie Universit\"{a}t Berlin, Arnimallee 14, 14195 Berlin, Germany}
\affiliation{Helmholtz-Zentrum f\"{u}r Materialien und Energie, Hahn-Meitner-Platz 1, 14109 Berlin, Germany}
\author{Aishwarya Chauhan}
\email{ph18d022@smail.iitm.ac.in}
\affiliation{Department of Physics, Indian Institute of Technology Madras, Chennai 600036, India}
\author{Yasir Iqbal}
\email{yiqbal@physics.iitm.ac.in}
\affiliation{Department of Physics, Indian Institute of Technology Madras, Chennai 600036, India}
\author{Johannes Reuther}
\email{reuther@zedat.fu-berlin.de}
\affiliation{Dahlem Center for Complex Quantum Systems and Institut f\"ur Theoretische Physik, Freie Universit\"{a}t Berlin, Arnimallee 14, 14195 Berlin, Germany}
\affiliation{Helmholtz-Zentrum f\"{u}r Materialien und Energie, Hahn-Meitner-Platz 1, 14109 Berlin, Germany}
\date{\today}

\begin{abstract}
We perform extensive classifications of $\mathds{Z}_2$ quantum spin liquids on the simple cubic, body centered cubic, and face centered cubic lattices using a spin-rotation invariant fermionic projective symmetry group approach. Taking into account that all three lattices share the same point group $O_h$, we apply an efficient gauge where the classification for the simple cubic lattice can be partially carried over to the other two lattices. We identify hundreds of projective representations for each of the three lattices, however, when constructing short-range mean-field models for the fermionic partons (spinons) these phases collapse to only very few relevant cases. We self-consistently calculate the corresponding mean-field parameters for frustrated Heisenberg models on all three lattices with up to third neighbor spin interactions and discuss the spinon dispersions, ground state energies and dynamical spin structure factors. Our results indicate that phases with non-uniform spinon hopping or pairing amplitudes are energetically favored. An unusual situation is identified for the fcc lattice where the spinon dispersion minimizing the mean-field energy features a network of symmetry protected line-like zero modes in reciprocal space. We further discuss characteristic fingerprints of these phases in the dynamical spin structure factor which may help to identify and distinguish them in future numerical or experimental studies.  
\end{abstract}
\maketitle

%====================================================================
\section{Introduction}\label{Intro}
In recent years, quantum spin liquids have become one of the most vibrant research fields in condensed matter physics~\cite{Balents-2010,Broholm-2020}. Besides the absence of magnetic order, these phases realize the fascinating scenario where long-range entanglement, topological order and fractional quasiparticle excitations combine to form novel quantum many-body states~\cite{Savary-2016,Zhou-2017}. Two main strategies of theoretical investigations are currently pursued: In a direct numerical treatment, a given spin Hamiltonian is investigated with respect to its magnetic correlations or excitations aiming to identify quantum spin liquid behavior. While this approach has led to invaluable insights into quantum spin liquids and possible Hamiltonians realizing them, powerful numerical methods are scarce and often limited by the general difficulty of probing topological order and fractional quasiparticles. The second strategy of approaching quantum spin liquids amounts to {\it proposing} effective low-energy theories for the system's fractional excitations which are then further theoretically studied. While within this strategy it is often difficult (if not impossible) to relate the considered theories to an actual spin Hamiltonian it allows for an investigation of quantum spin liquids on a fundamental level and in a systematic manner.

One approach related to this second strategy is the so-called projective symmetry group (PSG) method~\cite{wen02} which constitutes the central theme of this work. By reformulating the original spin degrees of freedom in terms of parton operators (which, here, are chosen to be fermionic)~\cite{Pomeranchuk-1941,baskaran87,baskaran88} the PSG approach allows one to classify possible free parton theories for quantum spin liquids based on the symmetries of the system. The partons may be identified with spinons (which are the fundamental spinful and fractional quasiparticle excitations of a quantum spin liquid) and via an additional coupling to an emergent gauge field the system may be conveniently described by a lattice gauge theory which is widely believed to capture the essential low-energy physics of a quantum spin liquid. Even though incapable of directly probing a given spin Hamiltonian with respect to a spin liquid ground state, a PSG classification may serve as a guide for further theoretical and experimental investigations. For example, the fermionic states obtained within a PSG analysis can be used as trial wave functions in a variational Monte Carlo study of specific quantum spin models~\cite{Capriotti-2001,Iqbal-2011,Hu-2013}. Furthermore, one may calculate dynamical spin structure factors for the classified spin liquid phases based on the two-parton excitation spectrum. Comparing these predictions with results from other numerical studies or neutron scattering experiments might allow one to identify and characterize spin liquid behavior for concrete spin Hamiltonians or even for real materials~\cite{Dodds-2013}. 

As the field of quantum spin liquids progresses and new systems beyond standard 2D spin models on triangular, honeycomb or kagome lattices are explored, the focus shifts more towards spin liquids in three dimensions (3D)~\cite{Balz-2016,Plumb-2019,Chillal-2020}. While quantum fluctuations generally decrease in higher dimensions, there is consensus that magnetic frustration can still be strong enough to melt magnetic long-range order. Since the numerical challenges of treating quantum spin systems increase further when going to 3D, analytical approaches such as the PSG become more important. However, there are so far only very few PSG studies classifying quantum spin liquids on 3D lattices~\cite{Huang-2017,Huang-2018,Liu-2019,Jin-2020}. Indeed, for the simple cubic (sc), body-centered cubic (bcc) and face-centered cubic (fcc) lattices representing classic textbook examples of 3D lattices, a PSG classification has not been achieved so far, even though the corresponding Heisenberg models are known for their rich quantum phase diagrams potentially hosting quantum paramagnetic states. For example, the antiferromagnetic $J_1$-$J_2$-$J_3$ Heisenberg model on the sc lattice, besides various commensurate magnetically ordered phases, has been proposed to host an extended non-magnetic regime in the vicinity of a classical triple point (possibly realizing a quantum spin liquid)~\cite{Laubach-2015,Iqbal-2016,Oitmaa-2017,Hu-2018}. Similarly, the antiferromagnetic $J_1$-$J_2$-$J_3$ Heisenberg model on the bcc lattice shows an interplay of five different magnetically ordered phases, including incommensurate spirals, where there is evidence that quantum fluctuations can melt the magnetic long-range order in certain parameter regimes~\cite{Ghosh-2019}. For the fcc lattice, already a nearest neighbor antiferromagnetic Heisenberg coupling frustrates the system and leads to a subextensive manifold of degenerate classical ground states forming lines in momentum space~\cite{Haar-1962,Smart-1966}. By adding a second neighbor coupling $J_2=J_1/2$ the classical ground state degeneracy is enhanced even more and manifests as surfaces in momentum space~\cite{Balla-2019}. In both cases, the classical degeneracies are expected to amplify quantum fluctuations promising a rich physical behavior when the spin magnitude is lowered towards the quantum limit $S=1/2$.

Also from a material perspective these lattices open up new directions of investigations. Mott insulating materials featuring $S=1/2$ magnetic moments and realizing cubic crystal systems have recently shown potential as candidates hosting the quantum spin liquid state or being proximate to one. In particular, the garnet compound Ca$_3$Cu$_2$GeV$_2$O$_{12}$ features $S=1/2$ Cu$^{2+}$ ions occupying the B-sites which realize a bcc lattice~\cite{Lussier-2019}. Neutron diffraction experiments find an absence of magnetic ordering down to $70$ mK and indicate a large frustration ratio of at least $f=13.29$. This behavior has been argued to originate from the likely proximity of this system to the quantum phase transition point in the $S=1/2$ $J_{1}$\textendash$J_{2}$ antiferromagnetic model, which is known to be at $J_{2}/J_{1}\sim0.7$~\cite{Schmidt-2002,Oitmaa-2004,Majumdar-2009,Pantic-2014,Farnell-2016,Ghosh-2019}. Recently, a double perovskite compound Ba$_2$CeIrO$_6$ has been argued to be an excellent realization of a pseudospin $j=1/2$ spin-orbit coupled Mott insulator on the fcc lattice with a high degree of frustration $f\sim13$~\cite{revelli19}. Although the system undergoes magnetic ordering argued to be driven by Kitaev interactions, an estimate of the exchange parameters places it in proximity to a putative quantum spin liquid phase of the $J_{1}$\textendash$J_{2}$ Heisenberg model. Another interesting $S=1/2$ fcc antiferromagnet that is the molecular antiferromagnet Cs$_{3}$C$_{60}$, wherein specific heat measurements have revealed the occurrence of both long-range antiferromagnetic order and a quantum paramagnetic state below $2.2$ K~\cite{Kasahara-2014}.

The results of our extensive PSG classifications can be summarized as follows: The fact that all three lattices share the same point group $O_h$ simplifies the calculation significantly. Particularly, we present a scheme that allows us to reuse the PSGs from the sc case when treating the other two lattices. Due to the large number of point group elements ($O_h$ maximizes their number in 3D) we obtain a plethora of PSGs with a $\mathds{Z}_2$ gauge structure, reaching several hundreds or even more than a thousand phases. However, when constructing actual parton mean-field theories for these PSGs, consisting of short-range hopping and pairing terms, the symmetries act as constraints and thus only very few relevant cases remain. Besides the most simple mean-field phases where hopping and pairing amplitudes are uniform on bonds of the same type, we identify cases where these terms show non-trivial sign structures or a special symmetry-induced locking between hopping and pairing. We further compare the mean-field energies for all relevant phases. While on a mean-field level, the ground state energies are certainly not accurate in terms of absolute numbers and would be significantly lowered when performing a more elaborate Gutzwiller projection, they still allow for a relative comparison between different phases. A rather general observation is that non-uniform mean-field models tend to have lower energies compared to the uniform ones. An interesting situation occurs for the fcc lattice where the energetically preferred parton state exhibits an unusual symmetry protected network of line-like zero modes in momentum space. Finally, we compare the dynamical spin structure factors of several mean-field phases and discuss characteristic patterns of response which in the future may serve as a guide to identify these phases in numerical or experimental studies.

The rest of the paper is organized as follows: We start with a general introduction into the PSG method in Sec.~\ref{PSG}. In the following Sec.~\ref{representations}, we outline the PSG classification for the sc, bcc, and fcc lattices more specifically. Afterwards, in Sec.~\ref{sr_example}, we demonstrate, as an example, the derivation of short-range mean-field models for the bcc lattice. The main results of our work are presented in Sec.~\ref{sr_mf_states} where we discuss in detail the relevant short-range mean-field states including their spinon dispersions, ground state energies and dynamical spin structure factors for all three lattices. The paper ends in Sec.~\ref{conclusion} with a discussion and conclusion. More explicit calculations and tables presenting details on the PSG classifications are contained in several appendices.

\section{General projective symmetry group approach}\label{PSG}
In this section we provide a general introduction into the projective symmetry group (PSG) approach which allows us to classify effective low-energy theories for quantum spin liquids based on their behavior under symmetry transformations. Our starting point is a general Heisenberg Hamiltonian on an arbitrary lattice,
\begin{equation}
H=\sum_{\mathbf{rr'}}J_{\mathbf{rr'}}\mathbf{S}_\mathbf{r}\cdot\mathbf{S}_\mathbf{r'}\;.\label{ham}
\end{equation}
The fermionic version of the PSG approach which we apply in the following first amounts to rewriting the spin operators in terms of fermionic parton operators $f_{{\mathbf{r}\alpha}}$ on each lattice site ${\mathbf r}$~\cite{Abrikosov-1965},
\begin{equation}
S^\mu_\mathbf{r} = \frac{1}{2}\sum_{\alpha \beta}f^\dagger_{\mathbf{r}\alpha}\tau^\mu_{\alpha\beta}f_{\mathbf{r} \beta}\label{abrikosov}
\end{equation}
where $\alpha=\uparrow,\downarrow$ and $\tau ^\mu$ ($\mu=x,y,z$) are the Pauli matrices. The parton operators may be naturally identified as the spinfull and fractional quasiparticle degrees of freedom of quantum spin liquids, called spinons. In Eq.~(\ref{abrikosov}) their fractional nature is directly expressed by the fact that one spin operator is decomposed into two partons.

The key property of the mapping onto a fermionic system via Eq.~(\ref{abrikosov}) is that it enlarges the Hilbert space. While the original spin model only corresponds to single fermionic occupancies on each site, the Hilbert space of the full fermion model also includes doubly occupied and vacant sites. This property might first appear as an obstacle since the physical content of any fermionic wave function is only obtained after Gutzwiller projection onto the singly occupied subspace. On the other hand, the parton representation has the advantage that it is directly associated with a local $SU(2)$ gauge freedom~\cite{baskaran87,baskaran88,affleck88,Dagotto-1988} (see below) and, hence, allows us to describe the system by an effective gauge theory, which is known to be central for the understanding of quantum spin liquids. In a zeroth order approximation, the gauge fields may be treated as (static) numbers which is equivalent to a standard mean-field decoupling of the quartic terms in the fermionic version of Eq.~(\ref{ham}). Neglecting magnetic contributions of the form $\sim \left\langle \mathbf{S}_\mathbf{r} \right\rangle\cdot \mathbf{S}_\mathbf{r'}$ (which are irrelevant for our description of quantum spin liquids) and performing the decoupling in the fermionic hopping and pairing channels,
\begin{equation}
\chi_{\mathbf{rr'}} \delta_{\alpha \beta} = 2\left\langle f^\dagger_{\mathbf{r}\alpha} f_{\mathbf{r'}\beta}\right\rangle\;,\;
\Delta_{\mathbf{rr'}} \epsilon_{\alpha \beta} =~-2\left\langle f_{\mathbf{r}\alpha} f_{\mathbf{r'}\beta} \right\rangle \label{eq:decoupling}
\end{equation}
the fermionic Hamiltonian becomes
\begin{align}\label{eq:fermion_hamiltonian}
H_\text{mf}=& \sum_{\left\langle \mathbf{rr'} \right\rangle} -\frac{3}{8} J_{\mathbf{rr'}}\left( \psi^\dagger_\mathbf{r} u_{\mathbf{rr'}} \psi_\mathbf{r'} + h.c. - \frac{1}{2}\text{Tr} \left[ u^\dagger_{\mathbf{rr'}} u_{\mathbf{rr'}}\right] \right)\notag\\ 
+& \sum_\mathbf{r} \psi^\dagger_\mathbf{r} a_\mu (\mathbf{r}) \tau^\mu \psi_\mathbf{r}\;.
\end{align}
Here, we have introduced the spinor fields $\psi^\dagger_\mathbf{r} = (f^\dagger_{\mathbf{r} \uparrow}, f_{\mathbf{r} \downarrow})$ and the Lagrange multipliers $a_\mu (\mathbf{r})$ that enforce the single occupancy constraint on the mean-field level (i.e., on average),
\begin{equation}\label{eq:one_p_constraint}
\left\langle \sum_\alpha f^\dagger_{\mathbf{r}\alpha}f_{\mathbf{r}\alpha} \right\rangle = 1\;,\;
\left\langle f^\dagger_{\mathbf{r}\alpha}f^\dagger_{\mathbf{r}\beta} \right\rangle = \left\langle f_{\mathbf{r}\alpha}f_{\mathbf{r}\beta} \right\rangle = 0 \quad \forall~\mathbf{r}\;.
\end{equation}
Note that the second condition is a consequence of the first one. The $2\times2$ matrix $u_{\mathbf{rr'}}$ contains the hopping ($\chi_{\mathbf{rr'}}$) and pairing ($\Delta_{\mathbf{rr'}}$) mean-field amplitudes and is often refered to as \textit{ansatz},
\begin{align}\label{eq:mean-field_matrix}
u_{\mathbf{rr'}} = \begin{pmatrix}
\chi^\dagger_{\mathbf{rr'}} & \Delta_{\mathbf{rr'}} \\
\Delta^\dagger_{\mathbf{rr'}} & -\chi_{\mathbf{rr'}}
\end{pmatrix}=i\alpha^0_{\mathbf{rr'}} \tau^0 + \alpha_{\mathbf{rr'}}^\mu \tau^\mu\;.
\end{align}
In this equation we have also expressed $u_{\mathbf{rr'}}$ in terms of Pauli matrices and the identity matrix $\tau^0$ where $\alpha_{\mathbf{rr'}}^0$ and $\alpha_{\mathbf{rr'}}^\mu$ are real coefficients. This representation will later become very useful.

The mean-field Hamiltonian only contains free fermion terms and can be readily solved, but the assumption of static fields $u_{\mathbf{rr'}}$ is uncontrolled and the resulting mean-field solution does not even describe a physical spin system. However, a proper low energy theory beyond mean-field can be obtained by reintroducing fluctuations around a self-consistently obtained saddle-point solution for $u_{\mathbf{rr'}}$, restoring an effective lattice gauge theory~\cite{kogut79}. Depending on whether these fluctuations act as variations of the overall {\it sign} of $u_{\mathbf{rr'}}$ or of the overall {\it complex phase} of $u_{\mathbf{rr'}}$, the resulting gauge theories are of $\mathds{Z}_2$ or $U(1)$ type which fundamentally characterizes the quantum spin liquids they describe. By construction, these effective gauge theories are strongly interacting where fermionic spinons (partons) couple to an emergent gauge field (whose excitations are referred to as visons) and, therefore, cannot be easily solved. The purpose of this work is not to study the actual gauge theories but to classify all possible mean-field Hamiltonians of the form of Eq.~(\ref{eq:fermion_hamiltonian}). Still, on a pure mean-field level, the {\it invariant gauge group} (IGG) which will be introduced below allows one to infer the type of gauge fluctuations ($SU(2)$, $U(1)$, or $\mathds{Z}_2$)~\cite{Wen-1990,wen91,wen02} that would arise, given an ansatz $u_{\mathbf{rr'}}$. We will initially assume a $\mathds{Z}_2$ gauge group since these simplest and most restricted types of gauge fluctuations yield gapped vison excitations which ensures stability of the theory beyond mean-field. However, when investigating short-range ans\"atze $u_{\mathbf{rr'}}$ we will still encounter situations where the gauge group is lifted to $U(1)$ or $SU(2)$. 

We now describe the PSG procedure of classifying $\mathds{Z}_2$ mean-field ans\"atze by exploiting the system's lattice symmetries. As mentioned before, the fermionic representation in Eq.~(\ref{abrikosov}) has a local $SU(2)$ gauge invariance which manifests in the freedom to perform gauge transformations $\psi_\mathbf{r} \rightarrow W_\mathbf{r} \psi_\mathbf{r}$ where $W_\mathbf{r}$ is an arbitrary site-dependent $2\times2$ $SU(2)$ matrix. In terms of the local fermionic basis states, this transformation acts as a rotation in the unphysical subspace of doubly occupied and vacant sites but keeps the physical spin states in the singly occupied subspace unchanged. Alternatively, one can implement a gauge transformation as an operation acting on the ansatz and not on the spinor,
\begin{align}
u_{\mathbf{rr'}} \rightarrow W^\dagger_\mathbf{r} u_{\mathbf{rr'}} W_\mathbf{r'}\;.
\end{align}
A generic mean-field Hamiltonian breaks the local $SU(2)$ gauge freedom of the original fermionic system. However, there still exists a subgroup $\mathcal{G} \subseteq SU(2)$ (which is at least $\mathds{Z}_2$) such that the ansatz remains invariant for all sites,
\begin{align}
u_{\mathbf{rr'}} = W^\dagger_\mathbf{r} u_{\mathbf{rr'}} W_\mathbf{r'}, \quad W_\mathbf{r} \in \mathcal{G}\;.
\end{align}

The basic idea behind the PSG is that due to the system's gauge invariance any symmetry operation may be combined with a gauge transformation,
\begin{align}
u_{\mathbf{rr'}} \rightarrow W^\dagger_{\mathcal{S}(\mathbf{r})} u_{\mathcal{S}(\mathbf{r}) \mathcal{S}(\mathbf{r'})} W_{\mathcal{S}(\mathbf{r'})}\;,
\end{align}
which is referred to as a {\it projective} implementation of symmetries. Here, $\mathcal{S}$ is an element of the system's symmetry group acting on the lattice sites. The condition that an ansatz $u_{\mathbf{rr'}}$ satisfies the projective implementation of $\mathcal{S}$ is then given by 
\begin{align}\label{eq:u_gauge}
G_{\mathcal{S}}^\dagger(\mathcal{S}(\mathbf{r})) u_{\mathcal{S}(\mathbf{r}) \mathcal{S}(\mathbf{r'})} G_{\mathcal{S}}(\mathcal{S}(\mathbf{r'}))= u_{\mathbf{r}\mathbf{r'}}\;.
\end{align}
Here, and in the following the specific site dependent gauge transformation which fulfills this equation is denoted by $G_{\mathcal{S}}({\mathbf{r}})$. In other words, even though an ansatz $u_{\mathbf{rr'}}$ seems to {\it naively} break the system's lattice symmetries there may still exist a suitable gauge transformation such that the generalized symmetry condition in Eq.~(\ref{eq:u_gauge}) is fulfilled. Different projective implementations $G_{\mathcal{S}}({\mathbf{r}})$ satisfying Eq.~(\ref{eq:u_gauge}), hence, allow one to distinguish between different spin liquid phases with the same physical symmetries~\cite{wen02}. The above may be summarized by noting that the PSG is an extension of the symmetry group (SG) by the IGG
\begin{align}\label{eq:psg_def}
\text{PSG} = \text{SG} \ltimes \text{IGG}\;.
\end{align}
The first purpose of this work is to classify all PSGs for systems with an octahedral point group using Eq.~(\ref{eq:u_gauge}). In a second step, we construct the corresponding ans\"atze $u_{\mathbf{rr'}}$ as self-consistent saddle-point solutions and discuss their properties such as spinon band structures and physically observable spin structure factors.

\section{PSG representations for cubic lattices}\label{representations}
We now apply the concepts outlined in the last section to derive the projective representations of symmetries for lattices with an octahedral point group. In the first Subsection~\ref{sec:sc} we start with the sc lattice, followed investigations of the bcc and fcc lattices in Subsections~\ref{sec:bcc} and~\ref{sec:fcc}, respectively. Particularly, we will demonstrate how the PSG classification of the sc lattice may be reused to treat the latter two systems.

\subsection{Simple cubic lattice}\label{sec:sc}
The point group of the sc lattice is the octahedral group $O_h$. One possible choice of defining its generators (which we apply throughout this work) is given by
\begin{align}\label{eq:point_group_generator}
& \Pi_{z}(x, y, z) = (-x, -y, z)\;,\notag \\
& \Pi_{y}(x, y, z) = (-x, y, -z)\;,\notag \\
& \Pi_{xy}(x, y, z) = (y, x, -z)\;,\notag \\
& I(x, y, z) = (-x, -y, -z)\;, \notag \\
& P(x, y, z) = (z, x, y)\;.
\end{align}
The full space group includes the translations
\begin{align}\label{eq:translation_generator}
& T_{x}(x, y, z) = (x + 1, y, z)\;,\notag \\
& T_{y}(x, y, z) = (x, y + 1, z)\;,\notag \\
& T_{z}(x, y, z) = (x, y, z + 1)\;,
\end{align}
where the components of ${\mathbf r}=(x,y,z)$ take integer values. For the bcc and fcc lattices considered below we will keep the convention that the lattice constant of the cubic unit cell is always set to unity.

Besides these lattice symmetries we assume that time-reversal symmetry is satisfied. While time reversal $\mathcal{T}$ does not change the lattice coordinates and commutes with all other symmetry operations it has a non-trivial action on the parton operators, $\mathcal{T}(f_{\mathbf{r} \uparrow}, f_{\mathbf{r} \downarrow}) = (f_{\mathbf{r} \downarrow}, -f_{\mathbf{r} \uparrow})$. It then follows that time reversal acts on the spinor fields as $\mathcal{T}(\psi_\mathbf{r}) = \left[(i\tau^2 \psi_\mathbf{r})^\dagger\right]^T$. It is convenient to perform a global gauge transformation $\psi_\mathbf{r} \rightarrow -i\tau^2 \psi_\mathbf{r}$ which yields a simplified action of time reversal: $\mathcal{T}(\psi_\mathbf{r}) = \left[(\psi_\mathbf{r})^\dagger\right]^T$. If we now implement $\mathcal{T}$ as an operation acting on the ansatz one finds $\mathcal{T}(u_{\mathbf{rr'}}) = -u_{\mathbf{rr'}}$ and likewise for the Lagrange multiplier fields $\mathcal{T}(a_\mu (\mathbf{r})) = -a_\mu (\mathbf{r})$.

A valid projective representation needs to obey the same algebraic relations as the system's space group itself. This yields a set of constraints on the representation. For example, all generators of the point group in Eq.~\eqref{eq:point_group_generator}, except for $P$ [which performs a rotation by $2\pi/3$ around the $(1,1,1)$-axis] map back onto the identity when applied twice. Thus they need to be represented by a cyclic group of order $2$ while $P$ forms a cyclic group of order $3$. Most importantly, the gauge transformation associated with the identity operation is the IGG, which in our case is $\mathds{Z}_2$. This means that in a projective construction the identity is only defined up to a sign factor. As demonstrated below,  different choices of these signs lead to different PSGs.  

To ensure that different representations are gauge inequivalent one has to fix the gauge. It is convenient to choose a gauge in which the gauge transformations $G_{T_\mu}(\mathbf{r})$ related to translations are represented by the identity matrix modulated with a spatial sign structure. As explained in Appendix~\ref{ap:gauge} one can find a gauge in which
\begin{align}\label{g_translations}
&G_{T_x}(\mathbf{r})=\eta_{z_x}^{z}\eta_{y_x}^{y}\tau^0\;, \notag \\
&G_{T_y}(\mathbf{r})=\eta_{z_y}^{z}\tau^0\;, \notag \\
&G_{T_z}(\mathbf{r})=\tau^0\;,
\end{align} 
where the signs $\eta_{z_x}=\pm1$, $\eta_{y_x}=\pm1$, and $\eta_{z_y}=\pm1$ can be chosen independently (at least if no other symmetries are considered). Hence, for a system with only translation symmetries $T_x$, $T_y$, $T_z$ one would find $2^3$ PSGs. Note that fixing the $G_{T_\mu (\mathbf{r})}$ matrices does not yet fix the entire gauge freedom but leaves the possibility to perform a global gauge transformation. The projective representations of the remaining point group generators and time-reversal are determined by considering successive applications of group transformations such that the combined operation is given by the identity. Using the fixed representation for $G_{T_\mu (\mathbf{r})}$ in Eq.~(\ref{g_translations}) one can show that the gauge transformations associated with the point group generators may be brought into the form $G_{\mathcal{S}}(\mathbf{r})=\eta_{\mathcal{S}}^{f_{\mathcal{S}}(\mathbf{r})} g_\mathcal{S}$ where $\eta_\mathcal{S}=\pm1$, $f_\mathcal{S}({\mathbf r})$ is a function yielding inter values for all sites ${\mathbf r}$, and $g_{\mathcal{S}}$ is a $2\times2$ $SU(2)$ matrix. An example of this procedure is given in Appendix~\ref{ap:gauge_example} where it is also demonstrated that as a result of the symmetry $P$ one finds $\eta_{z_x}=\eta_{y_x}=\eta_{z_y}\equiv\eta_X=\pm1$. All PSGs for the sc lattice are then given by
\begin{align}\label{eq:algebra}
&G_{T_z}(\mathbf{r})=\tau^0, \quad G_{T_y}(\mathbf{r})=\eta_X^{z}\tau^0, \quad G_{T_x}(\mathbf{r})=\eta_X^{z+y}\tau^0, \notag \\
&G_{\mathcal{T}}(\mathbf{r}) = \eta_{\mathcal{T}}^{x+y+z} g_\mathcal{T}, \quad g_\mathcal{T}^2 = \pm \tau^0, \notag \\
&G_{I}(\mathbf{r}) = \eta_{I}^{x+y+z} g_I, \quad g_I^2 = \pm \tau^0, \notag \\
&G_{\Pi_z}(\mathbf{r}) = \eta_{\Pi}^{x+y} g_{\Pi_z}, \quad g_{\Pi_z}^2 = \pm \tau^0, \notag \\
&G_{\Pi_y}(\mathbf{r}) = \eta_{\Pi}^{x+z} g_{\Pi_y}, \quad g_{\Pi_y}^2 = \pm \tau^0, \notag \\
&G_{\Pi_{xy}}(\mathbf{r}) = \eta_{X}^{xy}\eta_{{\Pi_{xy}}}^{z} g_{\Pi_{xy}}, \quad g_{\Pi_{xy}}^2 = \pm \tau^0, \notag \\
&G_{P}(\mathbf{r}) = \eta_{X}^{x(y+z)} \eta_{P}^{x+y} g_P, \quad g_P^3 = \pm \tau^0, \notag \\
&\left[g_\mathcal{T}, g_\mathcal{O} \right]_{\pm} =0, \quad \left[g_I, g_{\mathcal{O}\neq I} \right]_{\pm} =0, \quad  \left[g_{\Pi_z}, g_{\Pi_y} \right]_{\pm} =0, \notag \\
&g_{\Pi_z} g_{\Pi_{xy}} g_{\Pi_y}^{-1} g_{\Pi_{xy}}^{-1} g_{\Pi_y}=\pm \tau^0, \quad g_{\Pi_z} g_P g_{\Pi_y}^{-1} g_P^{-1} =\pm \tau^0, \notag \\
&g_P g_{\Pi_{xy}} g_P g_{\Pi_{xy}} g_{\Pi_y}^{-1} =\pm \tau^0, \quad \eta_{\Pi}\eta_{\Pi_{xy}}\eta_P=1.
\end{align}
where the generators of the point group are denoted by $\mathcal{O}$. All parameters $\eta_X$, $\eta_{\mathcal{T}}$, $\eta_{I}$, $\eta_{\Pi}$, $\eta_{{\Pi_{xy}}}$, and $\eta_{P}$ take the values $\pm1$ and $\left[ \ldots \right]_{\pm}$ stands for the commutator or anti-commutator.

It is worth emphasizing that Eq.~(\ref{eq:algebra}) has been obtained after performing a gauge transformation of the form $W(\mathbf{r}) = \eta^x_{w_x}\eta^y_{w_y}\eta^z_{w_z}\tau^0$ where $\eta_{w_x}=\pm1$, $\eta_{w_y}=\pm1$, $\eta_{w_z}=\pm1$. This gauge transformation acts on the projective representations of translations as $G_{T_\mu}(\mathbf{r})\rightarrow \eta_{w_\mu} G_{T_\mu}(\mathbf{r})$, yielding a global sign which can be absorbed by a redefinition of $G_{T_\mu}(\mathbf{r})$. Furthermore, the projective representations of the point group elements remain unaffected, except for $P$ and $\Pi_{xy}$. For these latter two symmetry operations the gauge transformation acts as $G_{\Pi_{xy}}(\mathbf{r})\rightarrow \eta^{x+y}_{w_x} \eta^{x+y}_{w_y} G_{\Pi_{xy}}(\mathbf{r})$ and $G_P(\mathbf{r})\rightarrow \eta^{x+z}_{w_x} \eta^{x+y}_{w_y} \eta^{y+z}_{w_z} G_P(\mathbf{r})$. Thus, by properly choosing $\eta_{w_\mu}$ one obtains the simplified sign structure of $G_{\Pi_{xy}}(\mathbf{r})$ and $G_P(\mathbf{r})$ as presented in Eq.~(\ref{eq:algebra}).

One finds that Eq.~(\ref{eq:algebra}) can be solved by $21$  gauge inequivalent sets of $g_\mathcal{S}$-matrices which are listed in Appendix~\ref{ap:irreps}. Note that in all these solutions one has $g_{\Pi_z}=g_{\Pi_y}=\tau^0$. The total number of combinatorially distinct PSGs is two to the power of the number of independent $\eta_\mathcal{S}$ parameters times the number of gauge inequivalent sets of $g_\mathcal{S}$ matrices. The condition $\eta_{\Pi}\eta_{\Pi_{xy}}\eta_P=1$ connects three different sign factors such that only two can be counted as independent. This yields  $21\cdot 2^5 = 672$ PSGs for the sc lattice. However, due to the property $\mathcal{T}(u_{\mathbf{rr'}}) = -u_{\mathbf{rr'}}$ it is clear that no finite mean-field ansatz can be constructed if projective time reversal acts trivially (i.e., $\eta_{\mathcal{T}}=1$ and $g_{\mathcal{T}}=\tau^0$). Hence, when investigating actual ans\"atze, only $21\cdot 2^5 -9\cdot 2^4 = 528$ cases need to be considered.

\subsection{Body centered cubic lattice}\label{sec:bcc}
We now extend the previous discussion to the bcc lattice. While the space group $O_h$ remains unaffected, a new generator for translations needs to be incorporated, which corresponds to a translation along the space diagonal by half the lattice constant of the cubic unit cell, 
\begin{align}\label{eq:bcc_translation}
t(x, y, z)=(x+1/2, y+1/2, z+1/2)\;.
\end{align}
By viewing the bcc lattice as two interpenetrating sc lattices with sublattice $A~=~\left\lbrace \mathbf{r}=(x,y,z)| x,y,z \in \mathbb{Z} \right\rbrace$ and $B~=~\left\lbrace\mathbf{r}=(x+1/2,y+1/2,z+1/2)|x,y,z \in \mathbb{Z} \right\rbrace$ we may reuse our results form the previous section. Here, we only sketch the procedure and refer to Appendix~\ref{ap:bcc} for details. Before including $t$, we assume that each of the two sublattices independently realizes one of the PSGs already classified. We may symbolically write this as $G_\mathcal{S}(\mathbf{r}\in A)=G_\mathcal{S}^A(\mathbf{r})$ and $G_\mathcal{S}(\mathbf{r}\in B)=G_\mathcal{S}^B(\mathbf{r})$ where $G_\mathcal{S}^{A/B}(\mathbf{r})$ fulfills Eq.~(\ref{eq:algebra}). Initially, this construction requires that the point group symmetries acting on sublattice $B$ need to leave one site $\mathbf{r}_0^B$ invariant in the same way as the point group symmetries leave the origin $\mathbf{r}_0^A=(0,0,0)$ on sublattice $A$ unchanged. We choose this site as $\mathbf{r}_0^B=(1/2, 1/2, 1/2)$. As an example, site inversion $I^B$ acting on sublattice $B$ does not obey $I^B(x+1/2,y+1/2,z+1/2)=(-x-1/2,-y-1/2,-z-1/2)$, as one would naively expect, but operates as $I^B(x+1/2,y+1/2,z+1/2)=(-x+1/2,-y+1/2,-z+1/2)$.

The extension by $t$, which connects the two sublattices, adds further algebraic conditions which are obtained from successive applications of symmetry operations yielding identity, similarly to the approach in the previous section. It can be shown that the representation matrices $g^{A}_{\mathcal{S}}=g^{B}_{\mathcal{S}}$ and the sign parameter $ \eta^{A}_{\mathcal{S}}=\eta^{B}_{\mathcal{S}}$ of the two sublattices have to be identical for all symmetries. An important consequence is that the sign factor corresponding to translations can only be positive $\eta^A_X=\eta^B_X=+1$. This also simplifies the handling of point group symmetries: Since inversion on sublattice $B$ obeys $I^B({\mathbf r}\in B)=T_xT_yT_zI({\mathbf r}\in B)$, where $I$ is the conventional inversion satisfying $I({\mathbf r})=-{\mathbf r}$ on both sublattices and $T_\mu$ is associated with a trivial gauge transformation, one finds that $G_I(\mathbf{r}\in B)=G_{I^B}(\mathbf{r}\in B)$. The same also holds for the other point group symmetries, such that one can implement them in the usual way where their action only leaves one point $\mathbf{r}_0=(0,0,0)$ invariant. In total, the gauge transformations associated with the symmetry operations are given by the same equations as for the sc lattice [Eq.~\eqref{eq:algebra}] but additional conditions for the projective representation of $t$ have to be included:
\begin{align}\label{eq:bcc_algebra}
& G_{t}(\mathbf{r})=\eta^{x + y + z}_{t} g_t, \quad g_t^2 = \pm \tau^0\;, \notag \\
& \left[g_{t}, g_{\mathcal{S}\neq\Pi_y,\Pi_z} \right]_{\pm} =0, \quad g_{\Pi_{xy}}g_Pg_tg_{\Pi_{xy}}g_P = \pm g_t\;.
\end{align}
Note that the last three identities hold because all translations $T_\mu$ are now represented by the identity and $g_{\Pi_z}=g_{\Pi_y}=\tau^0$. It is important to emphasize that the components $x$, $y$, $z$ in Eq.~(\ref{eq:bcc_algebra}) label the cubic unit cell of a site at position ${\mathbf r}$, i.e., for a site on sublattice $B$ they obey $\mathbf{r}=(x+1/2,y+1/2,z+1/2)$ with $x,y,z \in \mathbb{Z}$, see Fig.~\ref{fig:bcc_lattice}. The projective representations defined by the possible sets of $g_\mathcal{S}$ matrices are listed in Table~\ref{tab:reps_bcc}. Combined with the possible choices for the sign parameters, one obtains a total of $59\cdot 2^5=1888$ distinct PSGs for the body centered cubic. Subtracting again the cases where time reversal acts trivially such that no finite mean-field ansatz can be constructed, yields $59\cdot 2^5 - 23\cdot 2^4 = 1520$.

\subsection{Face centered cubic lattice}\label{sec:fcc}
We finally discuss the fcc lattice where we proceed in analogy to the bcc lattice. Compared to the sc case, one now has to add two more translations given by
\begin{align}\label{eq:fcc_translation}
& t_1(x, y, z) = (x, y + 1/2, z + 1/2)\;, \notag \\
& t_2(x, y, z) = (x + 1/2, y+1/2, z)\;.
\end{align}
The fcc lattice can be constructed by four sc sublattices defined by $A=\left\lbrace (x,y,z)| x,y,z \in \mathbb{Z} \right\rbrace$, $B=\left\lbrace (x+1/2,y+1/2,z)| x,y,z \in \mathbb{Z} \right\rbrace$, $C=\left\lbrace (x+1/2,y,z+1/2)| x,y,z \in \mathbb{Z} \right\rbrace$ and $D=\left\lbrace (x,y+1/2,z+1/2)| x,y,z \in \mathbb{Z} \right\rbrace$ which are connected by $t_1$ and $t_2$. Using the same line of arguments as for the bcc lattice we find that the gauge transformations again have to be represented equally on all sublattices, i.e., $G^A_{\mathcal{S}}=G^B_{\mathcal{S}}=G^C_{\mathcal{S}}=G^D_{\mathcal{S}}$. Furthermore, like for the bcc case, the sign factor corresponding to translations must be positive, $\eta_X = +1$ (which again simplifies the handling of point group symmetries due to the same reason already discussed for the bcc lattice). The gauge transformations associated with the new generators $t_1$ and $t_2$ and the additional algebraic relations for the $g_{t_1}$ and $g_{t_2}$ matrices have the form
\begin{align}\label{eq:fcc_algebra}
& G_{t_1}(\mathbf{r})=\eta_t^{x+y+z}g_{t_1} \quad g_{t_1}^2=\pm \tau^0, \notag \\
& G_{t_2}(\mathbf{r})=\eta_t^{x+y+z}g_{t_2} \quad g_{t_2}^2=\pm \tau^0, \notag \\
& \left(g_{t_1}g_{t_2}\right)^2=\pm \tau^0, \quad \left[g_\mathcal{T}, g_{t_1}\right]_{\pm} = 0, \quad  \left[g_\mathcal{T}, g_{t_2}\right]_{\pm} = 0, \notag \\
& \left[g_I, g_{t_1}\right]_{\pm} = 0, \quad  \left[g_I, g_{t_2}\right]_{\pm} = 0, \quad  \left[g_{\Pi_{xy}}, g_{t_2}\right]_{\pm} = 0, \notag \\
& g_{t_2}g_{t_1}g_{\Pi_{xy}}g_{t_1}g_{\Pi_{xy}}=\pm \tau^0, \quad g_Pg_{t_2}g_P^{-1}g_{t_1}= \pm \tau^0, \notag \\
& g_{\Pi_{xy}}g_Pg_{t_1}g_{\Pi_{xy}}g_Pg_{t_1} =\pm \tau^0.
\end{align}
We again emphasize that $x,y,z \in \mathbb{Z}$ are the coordinates of the cubic unit cell in which the site ${\mathbf r}$ lies. Note that there is only one sign factor $\eta_t$ for both transformations $t_1$ and $t_2$. Furthermore, in contrast to the bcc lattice one finds that Eq.~\eqref{eq:fcc_algebra} only allows for solutions where the matrix representations for the translations $t_1$ and $t_2$ are trivial, $G_{t_1}(\mathbf{r})=G_{t_2}(\mathbf{r})=\eta_t^{x+y+z} \tau^0$. As a consequence, one obtains the same gauge inequivalent sets of $g_{\mathcal S}$ matrices as for the sc lattice, see Table~\ref{tab:reps}. This means, the total number of PSGs is $21\cdot 2^5 = 672$ and after subtracting the ones where the gauge transformation of time reversal is trivial one finds $21\cdot 2^5 -9\cdot 2^4 = 528$.

\section{Constructing short-rang mean-field ans\"atze}\label{sr_example}
With the PSG representations at hand we are now able to construct mean-field ans\"atze which satisfy the projective symmetries. In this section, as an example, we explicitly construct such ans\"atze for the bcc lattice with nearest neighbor mean-field amplitudes. Afterwards, we will discuss ans\"atze for all three lattices with mean-field amplitudes up to third neighbors focussing more on their physical properties rather than their construction. Therefore, this section can be considered as a guide of how to use the PSG classification for constructing ans\"atze and readers only interested in the results may proceed to the next section.

\begin{figure}
\includegraphics[width=0.99\linewidth]{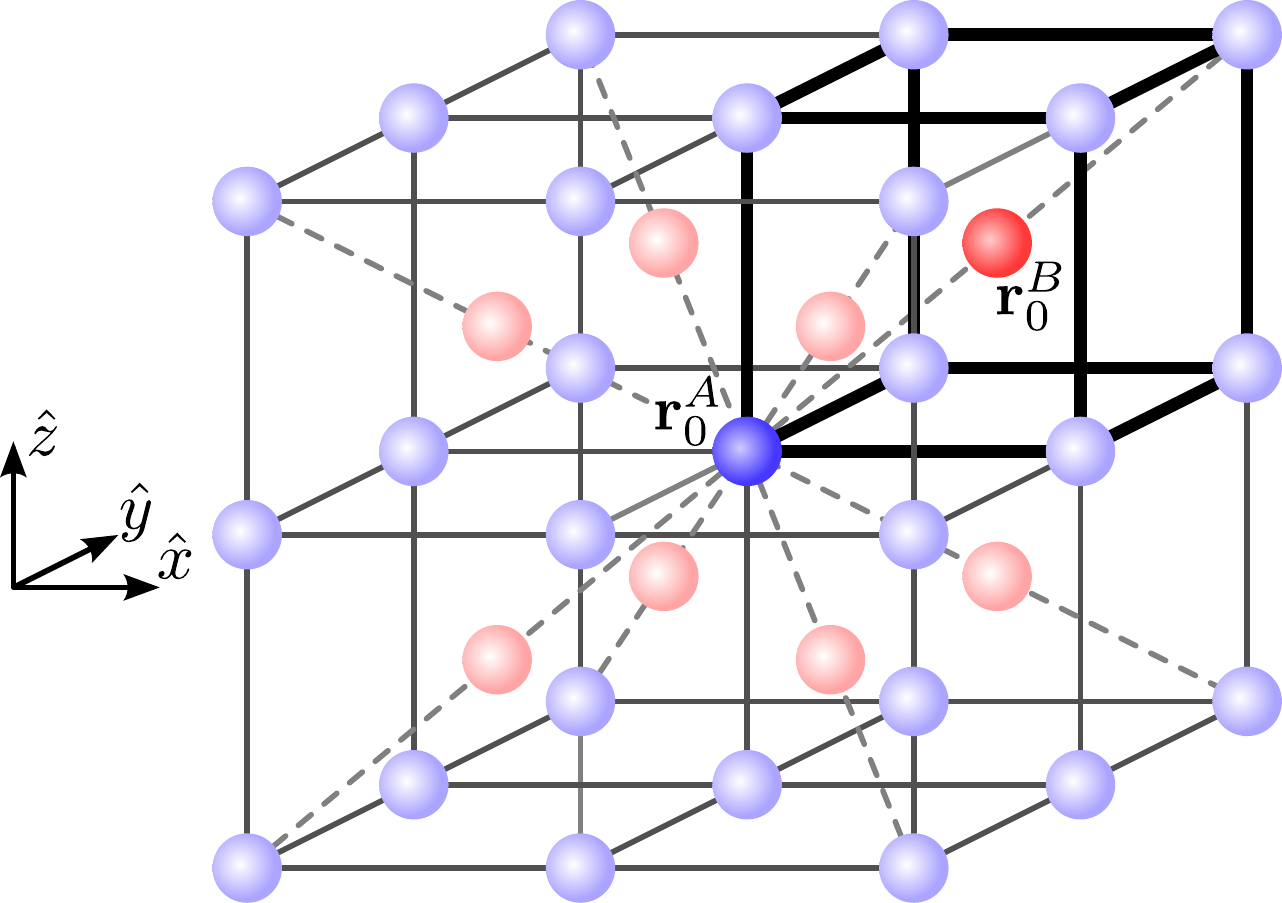}
\caption{Illustration of the bcc lattice where blue (red) points denote sublattice $A$ ($B$). The bold black lines in the upper right part of the figure highlight a cubic unit cell where the dark blue and dark red points are considered to lie inside this unit cell. The eight red points are the first neighbors of the dark blue site in the center.}\label{fig:bcc_lattice}
\end{figure}

The entire construction is based on Eq.~\eqref{eq:u_gauge} where the symmetry operators of the bcc lattice are given by $\mathcal{S}~=~\left\lbrace T_x, T_y, T_z, t, \mathcal{T}, I, \Pi_z, \Pi_y, \Pi_{xy}, P \right\rbrace$. Since the gauge transformations of translations are all represented by the identity $G_{T_\mu}(\mathbf{r})=\tau^0$, it immediately follows that
\begin{align}
u_{\mathbf{r}+\hat{e}_{\mu} \mathbf{r'}+\hat{e}_{\mu}} = u_{\mathbf{r r'}}\equiv u_{\delta{\mathbf r}}\;,
\end{align}
where $\delta \mathbf{r}=\mathbf{r'}-\mathbf{r}$. Note that this does not hold for the sc lattice where a negative sign factor $\eta_X=-1$ is possible. There are eight first neighbors on the bcc lattice described by the vectors $\delta \mathbf{r} = \left\lbrace \pm 1/2, \pm 1/2 , \pm 1/2 \right\rbrace$ where all combinations of signs are possible, as shown in Fig.~\ref{fig:bcc_lattice}.
Even though the mean-field matrices only depend on $\delta \mathbf{r}=\mathbf{r'}-\mathbf{r}$ and not on $\mathbf{r}$ and $\mathbf{r'}$ separately, we fix $\mathbf{r}=(0,0,0)$ as a reference point to simplify the discussion below. Thus, the nearest neighbor mean-field matrices considered here are $u_{\mathbf{r r'}}=u_{(0,0,0), (\pm1/2,\pm1/2,\pm1/2)}\equiv u_{(\pm1/2,\pm1/2,\pm1/2)}$. Among these matrices we can choose one, for instance $u_{(1/2, 1/2, 1/2)}\equiv u_{\delta \mathbf{r}_1}$, and all others follow by applying the point group operations. Before formulating relations between different $u_{(\pm1/2,\pm1/2,\pm1/2)}$, we first specify the general form of $u_{\delta \mathbf{r}_1}$. Time reversal dictates a property which has to be fulfilled by all $u_{\delta \mathbf{r}}$,
\begin{align}\label{eq:timerev_example}
& -G^\dagger_{\mathcal{T}}(\mathbf{r})u_{\mathbf{rr'}}G_{\mathcal{T}}(\mathbf{r'}) = u_{\mathbf{rr'}} \notag \\
\Longleftrightarrow\;& -\eta_{\mathcal{T}}^{x'+y'+z'} g^{-1}_{\mathcal{T}}u_{\delta \mathbf{r}}g_{\mathcal{T}}=u_{\delta \mathbf{r}}\;.
\end{align}
This means that for $\delta\mathbf{r}=\delta \mathbf{r}_1$ where $\mathbf{r}$ and $\mathbf{r'}$ lie in the same cubic unit cell the sign factor $\eta_{\mathcal{T}}$ cancels out. Therefore, $u_{\delta \mathbf{r}_1}$ has to anti-commute with the representation matrix $g_\mathcal{T}$ which is either given by $\tau^0$ or by $i\tau^2$ (see Table~\ref{tab:reps_bcc}). Since a finite matrix cannot anti-commute with the identity one finds $g_\mathcal{T}=i\tau^2$. This requires that in the expansion of the mean-field matrix two coefficients vanish, $\alpha^0_{\delta \mathbf{r}_1}=\alpha^2_{\delta \mathbf{r}_1}=0$, [see Eq.~\eqref{eq:mean-field_matrix}] and consequently $\left( u_{\delta \mathbf{r}_1} \right)^\dagger = u_{\delta \mathbf{r}_1}$. Generally, the effect of hermitian conjugation is  given by $\left( u_{\delta\mathbf{r}} \right)^\dagger = u_{-\delta \mathbf{r}}$ such that $u_{\delta \mathbf{r}_1}=u_{-\delta \mathbf{r}_1}=u_{(-1/2, -1/2, -1/2)}$. The vector $-\delta{\mathbf r}_1$ points from the origin to the cubic unit cell with the coordinates $(x, y, z)=(-1, -1, -1)$ such that the sign factor $\eta_{\mathcal{T}}$ does not cancel out in Eq.~\eqref{eq:timerev_example}. It is then obvious that only $\eta_{\mathcal{T}}=+1$ leads to a finite ansatz.

Combining hermitian conjugation and inversion leads to another condition that holds for all mean-field matrices,
\begin{align}\label{eq:herm_inv_example}
& G^\dagger_{I}(I(\mathbf{r}))u_{I(\mathbf{r})I(\mathbf{r'})}G_{I}(I(\mathbf{r'}))= u_{\mathbf{rr'}} \notag \\
\Longleftrightarrow\;& \eta_I^{I(x')+I(y')+I(z')}g^{-1}_I u_{-\delta \mathbf{r}}g_I=u_{\delta \mathbf{r}} \notag \\
\Longleftrightarrow\;& \eta_I^{I(x')+I(y')+I(z')}g^{-1}_I \left( u_{\delta \mathbf{r}}\right)^\dagger g_I=u_{\delta \mathbf{r}}\;.
\end{align}
In the case $\delta \mathbf{r}=\delta \mathbf{r}_1$ this condition demands that $u_{\delta \mathbf{r}_1}$ has to commute with the representation matrix $g_I$. Thus, $g_I$ can be represented by the identity or by $i\tau^3$. Since this equally holds for $u_{-\delta \mathbf{r}_1}$ one finds that the corresponding sign factor has to be positive, $\eta_I=+1$.

Next, we consider a requirement dictated by permutation: 
\begin{align}\label{eq:per_exmaple}
& G^\dagger_{P}(P(\mathbf{r}))u_{P(\mathbf{r})P(\mathbf{r'})}G_{P}(P(\mathbf{r'}))= u_{\mathbf{rr'}} \notag \\
\Longleftrightarrow\;& \eta_P^{P(x')+P(y')}g^{-1}_P u_{P(\delta \mathbf{r})}g_P=u_{\delta \mathbf{r}}\;.
\end{align}
Using $P(\delta \mathbf{r}_1)=\delta \mathbf{r}_1$ and observing that the sign factor cancels out it follows that $u_{\delta \mathbf{r}_1}$ has to commute with $g_P$ which can only be accomplished by a trivial representation $g_P=\tau^0$ (see Table~\ref{tab:reps_bcc}). In contrast to the considerations for time reversal $\mathcal{T}$ and inversion $I$, the case $\delta{\mathbf r}=-\delta \mathbf{r}_1$ does {\it not} lead to the condition $\eta_P=1$ in Eq.~(\ref{eq:per_exmaple}).

The other point group operations can be used to relate different $u_{(\pm1/2,\pm1/2,\pm1/2)}$ with each other:
\begin{align}
& \eta_\Pi^{\Pi_z(x')+\Pi_z(y')}u_{\Pi_z(\delta \mathbf{r})}=u_{\delta \mathbf{r}}\;, \\
& \eta_\Pi^{\Pi_y(x')+\Pi_y(z')}u_{\Pi_y(\delta \mathbf{r})}=u_{\delta \mathbf{r}}\;, \\
& \eta_{\Pi_{xy}}^{\Pi_{xy}(z')}g^{-1}_{\Pi_{xy}}u_{\Pi_{xy}(\delta \mathbf{r})}g_{\Pi_{xy}}=u_{\delta \mathbf{r}}.
\end{align}
Combining $\Pi_z$, $\Pi_y$, or $\Pi_{xy}$ with inversion leads to further conditions. For instance, one finds that $\Pi_{xy}(1/2, 1/2, 1/2)=(1/2, 1/2, -1/2)=\Pi_z(I(1/2, 1/2, 1/2))$ which yields $g_Ig^{-1}_{\Pi_{xy}}u_{\delta \mathbf{r}_1}g_{\Pi_{xy}}g_I^{-1}=u_{\delta \mathbf{r}_1}$, i.e., $g_{\Pi_{xy}}g_I^{-1}$ has to commute with $u_{\delta \mathbf{r}_1}$ and consequently $g_{\Pi_{xy}}=\tau^0$ or $g_{\Pi_{xy}}=i\tau^3$. Furthermore, from the relation $\Pi_{xy}(1/2, -1/2, -1/2)=I(1/2, -1/2, -1/2)=(-1/2, 1/2, 1/2)$ it follows that the sign factor for $\Pi_{xy}$ has to be positive, $\eta_{\Pi_{xy}}=+1$. The constraint $\eta_{\Pi_{xy}}\eta_{\Pi}\eta_P=1$ determines the remaining sign factor $\eta_P=\eta_\Pi$.

It remains to be shown how $t$ transforms the mean-field matrices. Using
\begin{align}
& G^{\dagger}_t(t(\mathbf{r}))u_{t(\mathbf{r})t(\mathbf{r'})}G_t(t(\mathbf{r'}))=u_{\mathbf{rr'}}
\end{align}
for the case $\mathbf{r'} - \mathbf{r}=\delta \mathbf{r}_1$ yields $\eta_t g^{-1}_{t} u_{\delta \mathbf{r}_1}g_t=u_{\delta \mathbf{r}_1}$ where the invariance of the mean-field matrices under lattice translations $T_x$, $T_y$, and $T_z$ was used. Repeating the same for $u_{-\delta \mathbf{r}_1}$ one finds $g^{-1}_{t} u_{\delta \mathbf{r}_1}g_t=u_{\delta \mathbf{r}_1}$. Thus, we conclude that $\eta_t=+1$ and $u_{\delta \mathbf{r}_1}$ has to commute with $g_t$ which leads to the two possibilities $g_t=\tau^0$ or $g_t=i\tau^3$.

Putting everything together we have identified all PSG representations on the nearest neighbor level which are distinguished by $\eta_P$ (which is either $+1$ or $-1$) and $g_I$, $g_{\Pi_{xy}}$, $g_t$ can all be independently given by $\tau^0$ or $i\tau^3$. One can subdivide these 16 PSGs into 2 groups (see Table~\ref{tab:psg_bcc_example}): In the first case $g_I=g_{\Pi_{xy}}=g_t=\tau^0$ and in the second case at least one of the matrices $g_I$, $g_{\Pi_{xy}}$, $g_t$ is given by $i\tau^3$. The latter representations (second line in Table~\ref{tab:psg_bcc_example}) require that an ansatz as given in Eq.~\eqref{eq:mean-field_matrix} has only finite $\alpha^3_{\delta \mathbf{r}}$ coefficients such that $u_{\delta \mathbf{r}}=\alpha^3_{\delta \mathbf{r}}\tau^3$ for all $\delta \mathbf{r}$ (i.e, not only for nearest neighbor distances). In the first case where $g_I=g_{\Pi_{xy}}=g_t=\tau^0$ the projective symmetries are less restrictive and an ansatz can have the general form $u_{\delta \mathbf{r}}=\alpha^1_{\delta \mathbf{r}}\tau^1 + \alpha^3_{\delta \mathbf{r}}\tau^3$. Particularly, the `direction' of an ansatz $u_{\delta \mathbf{r}}$ in the $\tau^1$-$\tau^3$-plane as defined by the coefficients $(\alpha^1_{\delta \mathbf{r}_1},\alpha^3_{\delta \mathbf{r}_1})$ is the same for all nearest neighbor $\delta\mathbf{r}$. Since all projective symmetries except for time-reversal are represented by the identity one can apply a global gauge transformation $W=e^{-i\theta \tau^2}$, with $\theta=\theta(\alpha^1_{\delta \mathbf{r}_1},\alpha^3_{\delta \mathbf{r}_1})$ denoting the polar angle in the plane spanned by $\tau^1$ and $\tau^3$, without altering the PSG representation. This gauge transformation rotates the nearest neighbor mean-field matrices along the $\tau^3$ axis and thus $\alpha^1_{\delta \mathbf{r}}=0$. After this rotation, there are only two distinct mean-field ans\"atze on the bcc lattice for nearest neighbor amplitudes which are distinguished by the sign parameter $\eta_P$. The precise form of these two ans\"atze and their physical properties are discussed in Sec.~\ref{sr_mf_states}.
\begin{table}
\setlength{\tabcolsep}{0.2cm}
\begin{tabular}{ >{$}c<{$}  >{$}c<{$}  >{$}c<{$}  >{$}c<{$}  >{$}c<{$}}
\hline\hline
\eta_{\mathcal{T}}g_\mathcal{T} & \eta_P g_P & \eta_I g_I & \eta_{\Pi_{xy}} g_{\Pi_{xy}} & \eta_t g_{t} \\
\hline
+i\tau^2  & \pm \tau^0 & +\tau^0 & +\tau^0 & +\tau^0 \\
+i\tau^2  & \pm \tau^0 & +\tau^0/+i\tau^3 & +\tau^0/+i\tau^3 & +\tau^0/+i\tau^3 \\
\hline\hline
\end{tabular}
\caption{Possible PSG representations for first neighbor ans\"atze on the bcc lattice. Note that in the second line at least one of the matrices $g_I$, $g_{\Pi_{xy}}$ or $g_t$ must be given by $i\tau^3$.}\label{tab:psg_bcc_example}
\end{table}

Some comments about the Lagrange multiplier fields are in order. In analogy to the relations for the mean-field matrices in Eq.~(\ref{eq:u_gauge}), they have to satisfy conditions ensuring the invariance under projective symmetries:
\begin{align}
G^\dagger_\mathcal{S}(S(\mathbf{r}))a_\mu(\mathcal{S}(\mathbf{r}))\tau^\mu G_\mathcal{S}(S(\mathbf{r}))=a_\mu(\mathbf{r})\tau^\mu\;.\label{condition_lagrange}
\end{align}
One immediately finds that $a_\mu(\mathbf{r}+\hat{e}_\nu)=a_\mu(\mathbf{r})\equiv a_\mu$ for $\nu=\left\lbrace x,y,z \right\rbrace$ by taking advantage of translational invariance. Since the two gauge transformations in Eq.~(\ref{condition_lagrange}) act on the same site, the $\eta$ factors square and, hence, become irrelevant. For the other symmetry operations the term $a_\mu \tau^\mu$ transforms according to
\begin{align}
-& g^{-1}_{\mathcal{T}}a_\mu \tau^\mu g_{\mathcal{T}}=a_\mu \tau^\mu\;, \notag \\
& g^{-1}_{\mathcal{O}}a_\mu \tau^\mu g_{\mathcal{O}}=a_\mu\tau^\mu\;,
\end{align} 
where $\mathcal{O}$ is a point group generator. In other words, $a_\mu \tau^\mu$ has to commute (anti-commute) with the representation matrix $g_\mathcal{O}$ ($g_{\mathcal{T}}$).

The above discussion shows that the matrix structure of $u_{\mathbf{rr'}}$ and the type of allowed Lagrange multipliers $a_\mu$, which both determine the mean-field Hamiltonian, are fixed by the PSG. However, symmetry properties alone do not determine the actual values of the nearest neighbor hopping amplitude $\alpha^3_{\delta \mathbf{r}_1} \equiv \chi_1$ and the chemical potential $a_3$. They may, however, be obtained self-consistently by calculating the expectation values in Eqs.~\eqref{eq:decoupling} and \eqref{eq:one_p_constraint} for the ground state of the mean-field Hamiltonian. These self-consistent mean-field theories form the basis for the discussions in the next section. 

\begin{figure*}
\includegraphics[width=\textwidth]{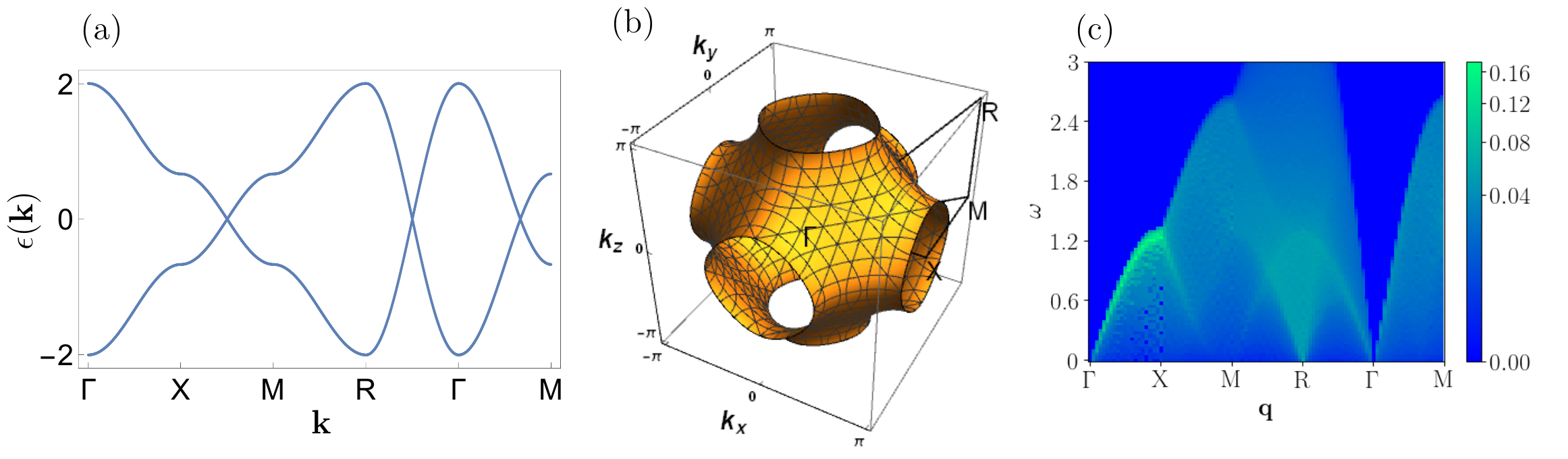}
\caption{Nearest neighbor model ``SC 1$_{1}$''. (a) Spinon dispersion of Eq.~\eqref{eq:sc_1st_ansatz} along the high-symmetry path through the Brillouin zone where $\Gamma=(0,0,0)$, $X=(\pi,0,0)$, $M=(\pi,\pi,0)$, and $R=(\pi,\pi,\pi)$ (and symmetry related wave vectors). The Fermi surface is depicted in (b). (c) Dynamical spin structure factor plotted along the high-symmetry path in reciprocal space.}\label{fig:sc_1_state}
\end{figure*}

\begin{table}[b]
\setlength{\tabcolsep}{0.2cm}
\begin{tabular}{ >{$}c<{$}  >{$}c<{$}  >{$}c<{$}  >{$}c<{$}  >{$}c<{$} }
\hline\hline
\eta_X& \eta_{\mathcal{T}}g_\mathcal{T} & \eta_P g_P & \eta_I g_I & \eta_{\Pi_{xy}} g_{\Pi_{xy}} \\
\hline
+ & +i\tau^2 & +\tau^0 & +\tau^0 & +\tau^0 \\
+ & +i\tau^2 & +\tau^0 & +\tau^0/+i\tau^3 & +\tau^0/+i\tau^3 \\
+ & -i\tau^3 & +\tau^0 & +\tau^0/-i\tau^2 & +\tau^0 \\
+ & -i\tau^3 & -\tau^0 & +\tau^0/-i\tau^2 & -i\tau^2\\
 - & +i\tau^2 & -\tau^0 & +\tau^0/+i\tau^3 & -i\tau^2 \\
- & -i\tau^3 & +\tau^0 & +\tau^0/-i\tau^2 & +i\tau^3 \\
\hline\hline
\end{tabular}
\caption{Possible PSG representations on the sc lattice which yield ans\"atze with symmetry-allowed first and second neighbor amplitudes. Note that some of the listed cases have been gauge transformed compared to Table~\ref{tab:reps} to ensure that the nearest neighbor ans\"atze all consist of hopping terms. Note that in the second line either $g_I$ or $g_{\Pi_{xy}}$ must be given by $i\tau^3$.}\label{tab:psg_sc_example}
\end{table}

\section{Short-range mean-field states}\label{sr_mf_states}
In Sec.~\ref{representations} we have shown that there exist hundreds of fermionic PSG representations for the sc, bcc, and fcc lattices. These large numbers follow from the fact that the octahedral group $O_h$ is the largest point group in three dimensions, containing a total of 48 elements. In simple terms, the larger the numbers of symmetries, the more algebraic relations between them exist, which increases the possibilities for constructing PSG representations. However, as demonstrated in the last section, when trying to determine actual mean-field ans\"atze with short range amplitudes only, the symmetries act as constraints which drastically reduce their number. Hence, the considered systems are characterized by a pronounced discrepancy between a large variety of PSGs but very limited numbers of mean-field theories, such that in this section only a few cases have to be discussed for each of the three lattices. This also implies that if quantum spin liquids exist in these systems their low energy effective theories and excitation spectra (e.g. spin structure factor) are already predetermined to some extent. This property possibly simplifies their identification in experiments. 

For each of the three lattices, we start with the nearest neighbor case and then add terms up to third neighbors. We emphasize that it is actually unlikely that a mean-field model with only nearest neighbor terms can describe a quantum spin liquid on the sc and bcc lattices~\cite{Kubo-1953,Marshall-1955,Kuzmin-2003b}. This is because on a mean-field level, the range of spinon hopping/pairing amplitudes is directly tied to the range of spin interactions $J_1,J_2,\ldots$ and beyond mean-field one may assume that such a constraint exists at least approximately. Therefore, one would expect that a nearest neighbor mean-field model only describes quantum spin liquids in systems with dominant nearest neighbor spin interactions $J_1$ in the presence of additional frustrated longer range interactions. However, without being frustrated, the sc system has been rigorously shown to order into a simple N\'eel state for $J_1>0$~\cite{Kennedy-1988} (where the two sublattices have opposite spin orientations) and, the same has been numerically demonstrated for the bcc lattice~\cite{Schmidt-2002} hence, a quantum spin liquid would not occur in these systems with nearest neighbor interactions only. We will still briefly consider this case, as it forms the basis for our investigations of longer-range models.

In the following, we discuss all the relevant cases for the three lattices.

\subsection{Simple cubic lattice}

On the sc lattice, two different types of mean-field ans\"atze can be constructed, and they are classified according to the sign value of $\eta_{X}$. The case of $\eta_{X}=+1$ corresponds to translationally invariant ans\"atze and $\eta_{X}=-1$ yields ans\"atze which double the unit cell in two of the three cubic lattice vector directions. We shall only consider mean-field ans\"atze with non-vanishing nearest neighbor amplitudes, and these correspond to PSG representations with $g_{P}=\tau_{0}$ in Table~\ref{tab:reps}.

\subsubsection{SC 1: $\eta_{X}=+1$ state}

This case is realized for the projective representations in the first four lines of Table~\ref{tab:psg_sc_example}. At the nearest neighbor level only a single ansatz with uniform hopping and a chemical potential can be constructed,
\begin{align}\label{eq:sc_1st_ansatz}
\text{SC 1$_1$}:  \quad &u_{\delta \mathbf{r}}=\chi_{1}\tau^3, \quad \forall~\delta \mathbf{r} \text{ first neighbors} \;,\notag \\
&a_3 \neq  0\;,
\end{align}
which realizes a gapless SU(2) spin liquid. Here, and in the following the notation ``SC X$_y$'' indicates the ansatz enumerated by ``X'' with ``$y$'' being the range of the mean-field amplitudes. Possible sub-cases for longer-range terms are labelled ``SC X$_{ya}$'', ``SC X$_{yb}$'', etc. The self-consistently calculated hopping amplitude $\chi_1$, on-site term $a_3$ and mean-field energy per site $\epsilon$ for $J_{1}=1$ are given by
\begin{equation}
\chi_1=0.167\;,\quad a_3=0.0\;,\quad \epsilon =-0.188\;.
\end{equation}
The spinon dispersion of this ansatz for both bands is shown in Fig.~\ref{fig:sc_1_state}(a). (Note that even though the dispersion of a uniform hopping term on a Bravais lattice can be presented with one band only, here and in the following, we prefer to use the two-component spinor basis to be consistent with cases where pairings are finite.) In Fig.~\ref{fig:sc_1_state}(b), we see the presence of a Fermi-surface which can be topologically characterized as a triply periodic Schwarz-P surface with an Euler characteristic $\chi=-4$~\cite{bubble_surfaces,Balla-2019}. The dynamical structure factor (see Appendix~\ref{ap:structure_factor} for a brief explanation of how the structure factor is calculated) shown in Fig.~\ref{fig:sc_1_state}(c) displays two principal variations in intensity, the first one is dispersive arising out of the $\Gamma$ point with strong and localized distribution of spectral weight at progressively higher $\omega$ as one traverses the $\overline{\Gamma X}$ segment. This feature is a direct consequence of the system's Fermi surface. The second noticeable characteristic is the appearance of a relatively weaker cone like signal around the R point. 

\begin{figure}[b]
\includegraphics[width=0.9\columnwidth]{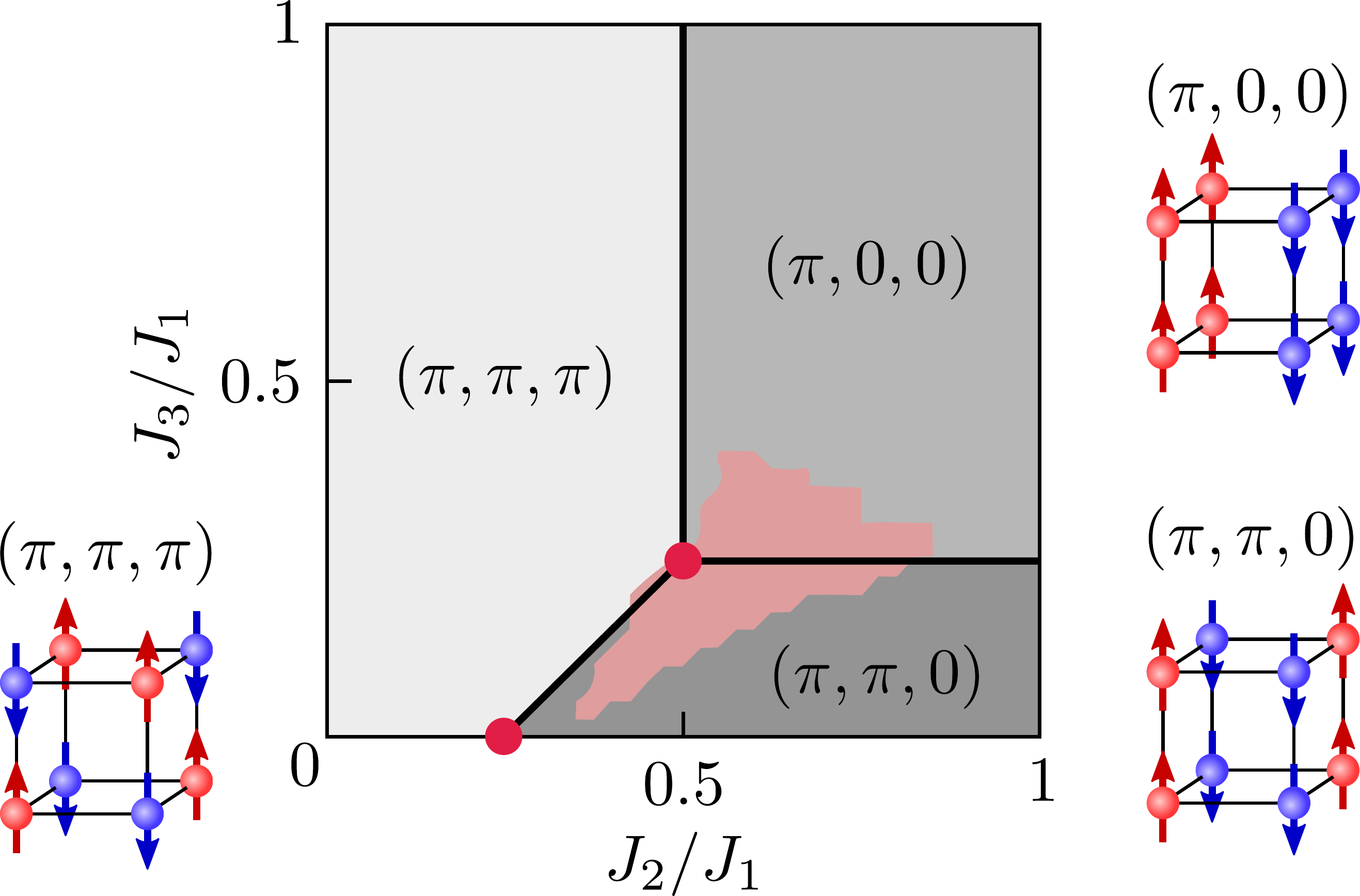}
\caption{Quantum phase diagram of the $S=1/2$ antiferromagnetic $J_1$-$J_2$-$J_3$ Heisenberg model on the sc lattice. Gray regions denote the classical phases with the corresponding ordering wave vectors indicated. Spin configurations are illustrated for all classical orders. Thick black lines are the classical phase boundaries. The red area is the regime where Ref.~\cite{Iqbal-2016} identifies a quantum paramagnetic phase. Red points mark the sets of Heisenberg couplings considered here.}\label{fig:sc_phase_diag}
\end{figure}

There exist three distinct ways of incorporating further neighbor amplitudes on top of the nearest neighbor ansatz of Eq.~\eqref{eq:sc_1st_ansatz}. The first and most general scenario corresponds to the PSG in the first line of Table~\ref{tab:psg_sc_example} which allows for the simultaneous occurrence of hopping and pairing amplitudes on second and third nearest neighbor bonds,
\begin{align}\label{eq:sc_2nd_ansatz_1}
&\text{SC 1$_{2a}$}: \quad u_{\delta \mathbf{r}}=\chi_{2}\tau^3 + \Delta_{2}\tau^{1},\; \forall~\delta \mathbf{r} \text{ second neighbors}\;,\notag\\
&\text{SC 1$_{3a}$}: \quad u_{\delta \mathbf{r}}=\chi_{3}\tau^3 + \Delta_{3}\tau^{1},\; \forall~\delta \mathbf{r} \text{ third neighbors}\;.
\end{align}
Here, second (third) neighbor bonds are of the form $\delta\mathbf{r}=(\pm1,\pm1,0)$ and permutations of coordinates ($\delta\mathbf{r}=(\pm1,\pm1,\pm1)$). Note that the second neighbor terms in Eq.~(\ref{eq:sc_2nd_ansatz_1}) lower the IGG down to $U(1)$, in particular, the $\Delta_{2}$ term opens a gap in the spinon spectrum except of nodal Dirac points along $\overline{\Gamma R}$ at $(\pi/2,\pi/2,\pm\pi/2)$. The inclusion of third neighbor terms further reduces the IGG down to $\mathds{Z}_2$.

The second way of including further neighbor amplitudes (``SC 1$_{2b}$'' and ``SC 1$_{3b}$'') is given by the second line of Table~\ref{tab:psg_sc_example}. Compared to Eq.~(\ref{eq:sc_2nd_ansatz_1}) the projective implementation of symmetries forbid spinon pairing terms, i.e., $\Delta_2=\Delta_3=0$. Our self-consistent calculations indicate that for a generic set of interaction parameters in the Hamiltonian, the $\Delta_2$ and $\Delta_3$ terms are finite and lower the mean-field energies $\epsilon$ such that the PSG in the first line turns out to be energetically favorable, in general. Therefore, we will not further discuss the case $\Delta_2=\Delta_3=0$, but instead focus on the more general type of ansatz in Eq.~(\ref{eq:sc_2nd_ansatz_1}).

The third way corresponds to the different cases in the third and fourth lines in Table~\ref{tab:psg_sc_example}. In this ansatz class, the projective symmetries dictate a uniform second neighbor imaginary pairing term and a third neighbor real hopping term,
\begin{align}\label{eq:sc_2nd_ansatz_2}
\text{SC 1$_{2c}$}: \quad u_{\delta \mathbf{r}}=&\Delta_{2}\tau^2, \quad \forall~\delta \mathbf{r} \text{ second neighbors}\;,\notag\\
\text{SC 1$_{3c}$}: \quad u_{\delta \mathbf{r}}=&\chi_{3}\tau^3, \quad \forall~\delta \mathbf{r} \text{ third neighbors}\;.
\end{align}
This case may, likewise, be obtained from the general ansatz in Eq.~(\ref{eq:sc_2nd_ansatz_1}) by setting $\chi_2=\Delta_3=0$ and performing a global gauge transformation around the $\tau^3$ axis (which, however, changes the $g$-matrices in the third and fourth lines in Table~\ref{tab:psg_sc_example}). Since the exclusion of $\chi_2$ and $\Delta_3$ terms again increases the energy this case also does not need to be considered separately.

We consider the extension in Eq.~(\ref{eq:sc_2nd_ansatz_1}) for two special coupling scenarios for $J_{2}$ and $J_{3}$ where enhanced quantum fluctuations are expected, thereby increasing the propensity for spin liquid behavior. The first scenario is given by $(J_{2}/J_{1},J_{3}/J_{1})=(0.25,0)$ where the corresponding classical model undergoes a phase transition between the $\mathbf{q}=(\pi,\pi,\pi)$ N\'eel and $\mathbf{q}=(\pi,\pi,0)$ stripe ordered phases [see Fig.~\ref{fig:sc_phase_diag}], and some studies have hinted at the possible existence of a nonmagnetic phase in the vicinity of this point for the $S=1/2$ model~\cite{Barabanov-1995,Majumdar-2010,Farnell-2016}. In the second scenario, we consider $(J_2/J_1,J_3/J_1)=(0.5,0.25)$ which is a triple point of the $\mathbf{q}=(\pi,\pi,\pi)$, $\mathbf{q}=(\pi,\pi,0)$ and $\mathbf{q}=(\pi,0,0)$ phases in the corresponding classical model [see Fig.~\ref{fig:sc_phase_diag}]. Recent studies~\cite{Iqbal-2016,Laubach-2015,Oitmaa-2017} have identified a nonmagnetic phase [marked by the red area in Fig.~\ref{fig:sc_phase_diag}] in the vicinity of this point for the $S=1/2$ model.

\begin{figure}
\includegraphics[width=\columnwidth]{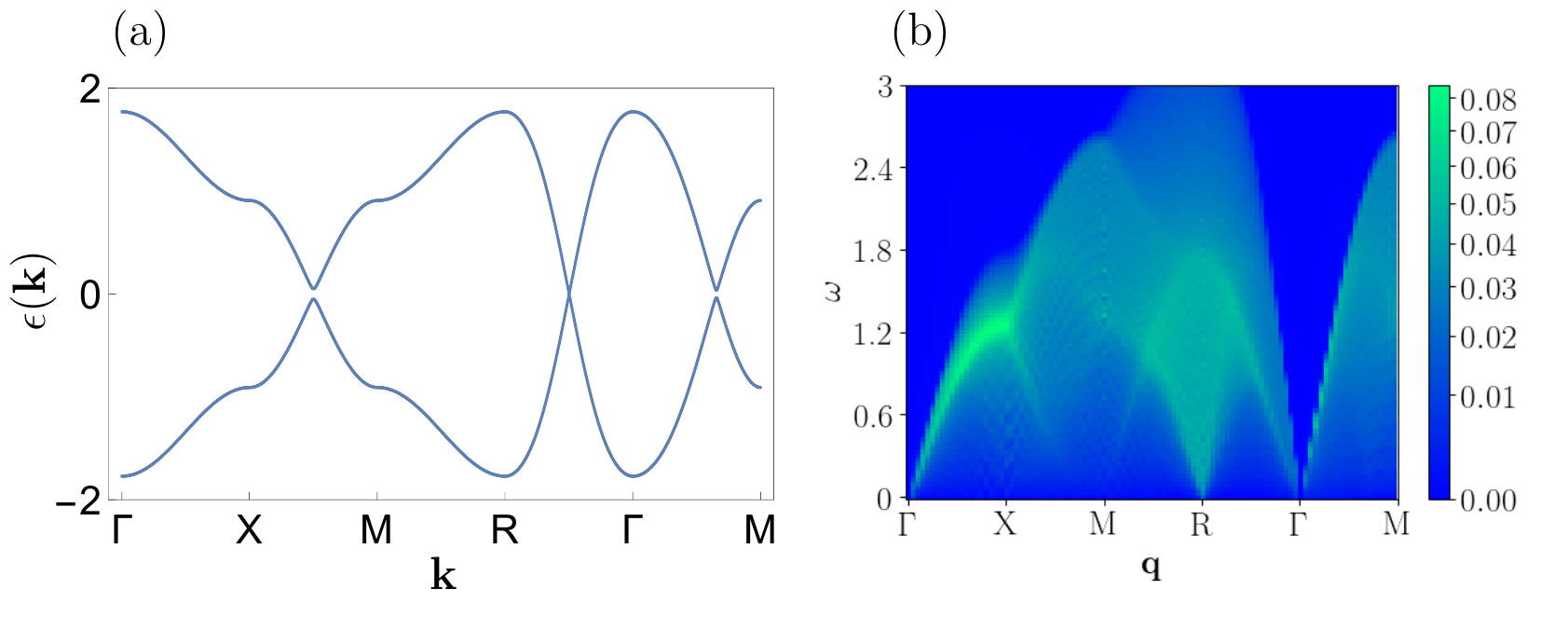}
\caption{``SC 1$_{3a}$'' model with mean-field amplitudes up to third neighbors. (a) Self-consistent spinon dispersion for the first neighbor terms in Eq.~\eqref{eq:sc_1st_ansatz}, second and third neighbor terms in Eqs.~\eqref{eq:sc_2nd_ansatz_1} (fixed to their self-consistently determined values given in Eq.~\eqref{eq:j1j2j3par}) along the high symmetry path through the first Brillouin zone. (b) Corresponding dynamical spin structure factor along the same path in reciprocal space.}
\label{fig:sc1_3a_state}
\end{figure}

For the first set of couplings $(J_{2}/J_{1},J_{3}/J_{1})=(0.25,0)$ the ansatz in Eqs.~(\ref{eq:sc_2nd_ansatz_1}) yields self-consistently calculated amplitudes given by 
\begin{align}
&\chi_1=0.167\;,\quad \chi_{2}=0.0\;,\quad \Delta_{2}=1.97\cdot10^{-3}\;,\notag \\
&a_3=0.0\;,\quad \epsilon =-0.188\;,\notag
\end{align}
which does not lead to any noticeable changes compared to the $J_1$ only case. In the second coupling scenario, at $(J_2/J_1,J_3/J_1)=(0.5,0.25)$, we find a small additional $\chi_{3}$ term and a comparatively smaller $\Delta_{2}$ term:  
\begin{align}\label{eq:j1j2j3par}
&\chi_1=0.167\;,\; \chi_{2}=0.0\;,\; \Delta_{2}=0.0127\;,\;\chi_3=-0.0598\;,\notag \\
&\Delta_{3}=0.0\;,\; a_{3}=0.0\;,\; \epsilon =-0.197\;.
\end{align}
As expected, the presence of a finite $\Delta_{2}$ in the self-consistent parameters of the SC 1$_{3a}$ ansatz [Eq.~\eqref{eq:j1j2j3par}], gaps out the Fermi surface leaving behind nodal Dirac points along $\overline{\Gamma R}$ at $(\pi/2,\pi/2,\pm\pi/2)$ [see Fig.~\ref{fig:sc1_3a_state}(a)]. Due to the smallness of $\Delta_{2}$ term, its manifestation in the dynamical spin structure factor is not visible, while, we notice that the effect of a finite $\chi_{3}$ is to suppress the intensity and broaden the relatively sharp signal [see Fig.~\ref{fig:sc1_3a_state}(b)] of the $\chi_{1}$ only case [Eq.~\eqref{eq:sc_1st_ansatz} and Fig.~\ref{fig:sc_1_state}(c)] along the $\overline{\Gamma X}$ segment.

\subsubsection{SC 2: $\eta_{X}=-1$ state}

\begin{figure}[t]
\includegraphics[width=\columnwidth]{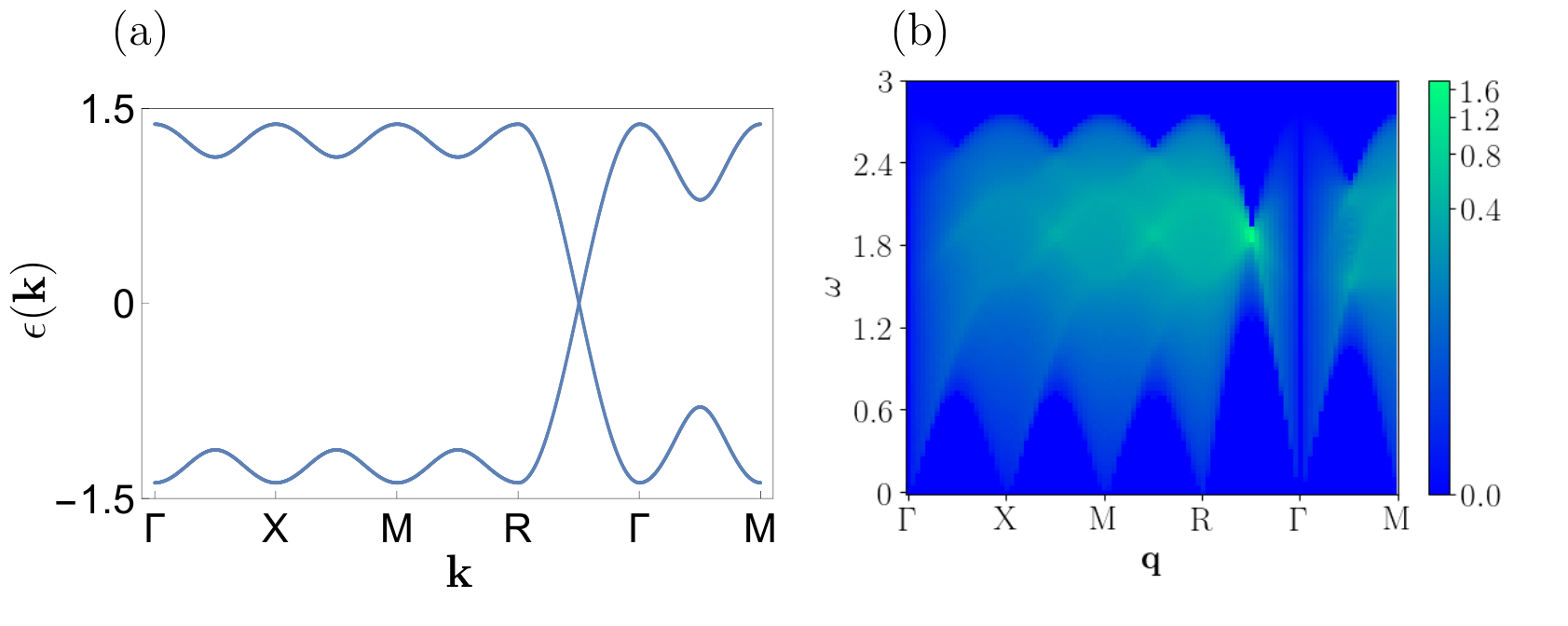}
\caption{Nearest neighbor model ``SC 2$_{1}$''. (a) Spinon dispersion of Eq.~\eqref{eq:sc2_1st_ansatz} along the high-symmetry path through the Brillouin zone. (b) Dynamical spin structure factor plotted along the high-symmetry path in reciprocal space.}
\label{fig:sc2_1_state}
\end{figure}

The mean-field ans\"atze in this case corresponding to the last two lines of Table~\ref{tab:psg_sc_example} require doubling the unit cell in both $x$- and $y$-directions. At the nearest neighbor level one obtains the following sign structure of real hopping terms
\begin{align}\label{eq:sc2_1st_ansatz}
\text{SC 2$_1$}:~&u_{(\pm1,0,0)}= \chi_{1}\tau^3\;, \notag \\
&u_{(0,\pm1,0)}= \eta_X^{x} \chi_{1}\tau^3\;, \notag \\
&u_{(0,0,\pm1)}= \eta_X^{(x + y)} \chi_{1}\tau^3, 
\end{align}
and a uniform onsite chemical potential term $a_{3}$. This ansatz realizes a $SU(2)$ spin liquid which is gapless at two isolated points $(\pi/2,\pi/2,\pm\pi/2)$ in the reduced Brillouin zone $k_{x}\in(0,\pi), k_{y}\in(0,\pi), k_{z}\in(-\pi,\pi)$. The self-consistently calculated hopping amplitude $\chi_1$, on-site term $a_3$ and mean-field energy per site $\epsilon$ for $J_{1}=1$ are given by
\begin{equation}\label{eq:sc2_1st_amplitudes}
\chi_1=0.199\;,\quad a_3=0.0\;,\quad \epsilon =-0.267\;.
\end{equation}
This energy is considerably lower compared to that of Eq.~\eqref{eq:sc_1st_ansatz}. The spinon dispersion of this state is shown in Fig.~\ref{fig:sc2_1_state}(a). The dynamical spin structure factor in Fig.~\ref{fig:sc2_1_state}(b) displays an entirely different distribution of signal compared to the SC 1 case with weakly dispersing features at low energies around the X, M and R points, while at intermediate energies one observes a high intensity concentration of diffuse spectral weight.

The inclusion of second neighbor amplitudes in the ansatz of Eq.~\eqref{eq:sc2_1st_ansatz} follows a similar scheme as in the SC 1 case. The most general second neighbor extension is given by the fifth line of Table~\ref{tab:psg_sc_example} when $\eta_{I}g_{I}=+\tau^{0}$, allowing for a simultaneous existence of hopping and pairing terms:
\begin{align}\label{eq:sc2_2nd_ansatz_2}
\text{SC 2$_{2a}$}:~&u_{(\pm1,\pm1,0)}=\eta_X^{x}(\chi_{2}\tau^3+\Delta_{2}\tau^{1})\;, \notag \\
&u_{(\pm1,0,\pm1)}=-\eta_X^{(x + y)} (\chi_{2}\tau^3+\Delta_{2}\tau^{1}) \;, \notag \\
&u_{(0,\pm1,\pm1)}=\eta_X^{y} (\chi_{2}\tau^3+\Delta_{2}\tau^{1})\;, \notag \\
&a_{3}=0\;.
\end{align}
Here, $(\pm1,\pm1,0)$ denotes the four bonds $(1,1,0)$, $(1,-1,0)$, $(-1,1,0)$, $(-1,-1,0)$ and equivalently for the other terms. This ansatz lowers the IGG from $SU(2)$ to $U(1)$, and splits the degeneracy of the bands but keeps the gapless point along $\overline{\Gamma R}$ intact. Other ways of including second neighbor terms such as for the case $\eta_{I}g_{I}=+i\tau^{3}$ in the fifth line or the various cases in the last line of Table~\ref{tab:psg_sc_example} are more restrictive and forbid parts of the terms in Eq.~(\ref{eq:sc2_2nd_ansatz_2}). In either case, however, self-consistently calculated second neighbor terms are vanishingly small at moderate $J_2$. Similarly, third neighbor terms are either forbidden by symmetry or numerically evaluate to very small values. Thus, this spin liquid phase is rather insensitive with respect to $J_2$ and $J_3$ couplings such that the self-consistent mean-field amplitudes, spinon dispersion and dynamical spin structure factor for both sets of spin interactions are again given by Eqs.~(\ref{eq:sc2_1st_ansatz}), (\ref{eq:sc2_1st_amplitudes}) and Fig.~\ref{fig:sc2_1_state}.

Since the mean-field energies of the SC 2 case are significantly lower compared to the SC 1 ansatz, we conclude that Fig.~\ref{fig:sc2_1_state}(b) represents a typical intensity distribution of the dynamical spin structure factor for possible quantum spin liquids on the sc lattice. Our analysis also shows that third neighbor amplitudes are required in the ans\"atze to realize a $\mathds{Z}_2$ quantum spin liquid on the sc lattice. A summary of the short-range mean field models and the corresponding projective implementations of symmetries is given in Table~\ref{tab:sc_summary}.

\subsection{Body centered cubic lattice}

\begin{figure*}
\includegraphics[width=\textwidth]{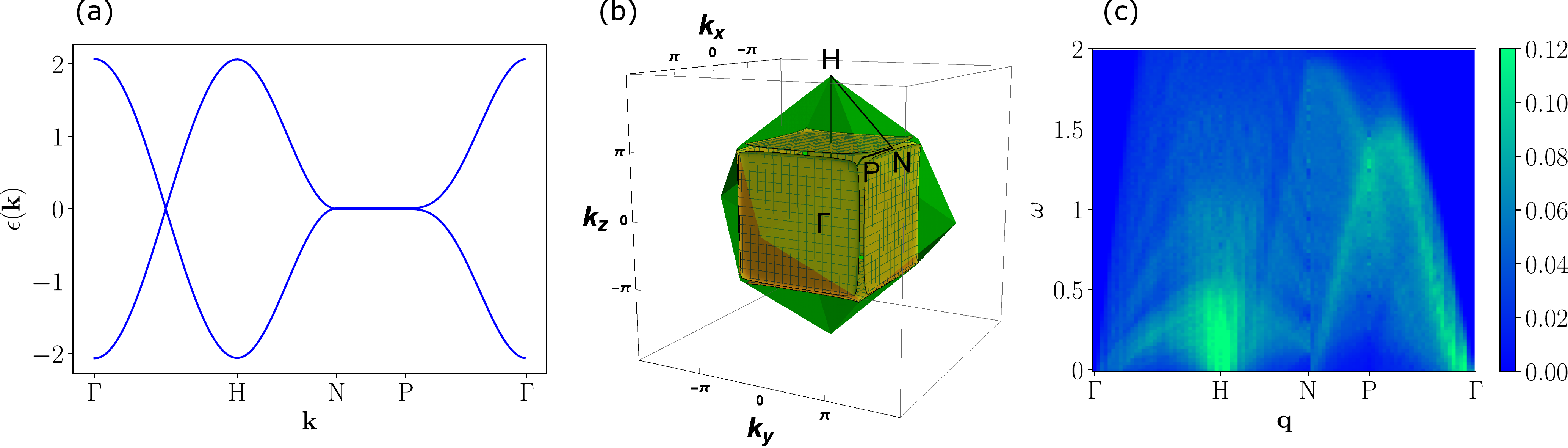}
\caption{Nearest neighbor model ``BCC 1$_1$''. (a) Spinon dispersion of Eq.~\eqref{eq:bcc_1a} along a path through the first Brillouin zone where $H=(0,0,2\pi)$, $N=(0,\pi,\pi)$, and $P=(\pi,\pi,\pi)$ (and symmetry related wave vectors). The Fermi surface is depicted in (b) where the green region indicates the first Brillouin zone. (c) Dynamical spin structure factor along a path in reciprocal space.}\label{fig:bcc_1_state}
\end{figure*}

We have already found in Sec.~\ref{sr_example} that on the nearest neighbor level, the bcc lattice only allows for two different ans\"atze which are distinguished by their sign factor $\eta_P$. While in the case $\eta_P=+1$  (referred to as BCC 1) only uniform hopping and pairing amplitudes are possible, the representations with  $\eta_P=-1$ (called BCC 2) are characterized by mean-field amplitudes which are modulated by certain sign patterns. In the following two subsections we discuss these cases in more detail and demonstrate how they can be physically distinguished by their spin structure factor. 

\subsubsection{BCC 1: $\eta_P = +1$ state}

The BCC 1 mean-field Hamiltonian on the nearest neighbor level only contains a simple uniform hopping term und a chemical potential
\begin{align}\label{eq:bcc_1a}
\text{BCC 1$_1$}: \quad &u_{\delta \mathbf{r}}=\chi_{1}\tau^3, \quad \forall~\delta \mathbf{r} \text{ first neighbors} \;,\notag \\
&a_3 \neq  0\;,
\end{align}
for which the IGG is $SU(2)$. The self-consistently calculated hopping amplitude $\chi_1$, on-site term $a_3$ and mean-field energy per site $\epsilon$ for $J_{1}=1$ are given by
\begin{equation}
\chi_1=0.129\;,\quad a_3=0.0025\;,\quad \epsilon =-0.149\;.
\end{equation}
The spinon dispersion of this ansatz for both bands is shown in Fig.~\ref{fig:bcc_1_state}(a). The system exhibits a Fermi surface, illustrated in Fig.~\ref{fig:bcc_1_state}(b), which consists of (almost) parallel planes forming a cube in momentum space. Due to the presence of a small $a_3$ term, the Fermi surface is slightly distorted compared to a perfect cube. The dynamical structure factor illustrated in Fig.~\ref{fig:bcc_1_state}(c) shows strong intensities around the $H$ point (i.e., ${\mathbf q}=(2\pi,0,0)$ and symmetry related wave vectors). This spectral distribution can be understood from the form of the Fermi surface in which two opposite planes are connected by a nesting vector ${\mathbf q}=(2\pi,0,0)$. A second characteristic is the cone-like signal around the $\Gamma$ point. The opening angle of the cone can be linked to the spinon Fermi velocity $v_\text{F}$. Comparing this angle for different directions emanating from the $\Gamma$ point, one finds that it is smaller on the line $\overline{\Gamma P}$ than on the line $\overline{\Gamma H}$ indicating a momentum dependent Fermi velocity.
\begin{figure}[b]
\includegraphics[width=\columnwidth]{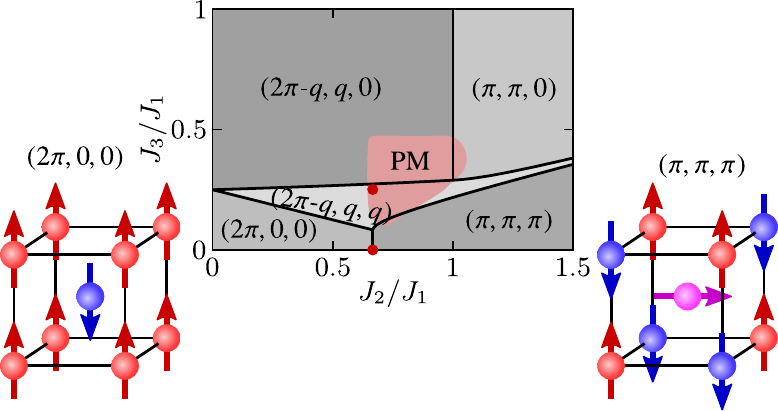}
\caption{Phase diagram of the classical antiferromagnetic $J_1$-$J_2$-$J_3$ Heisenberg model on the bcc lattice. Gray regions denote the classical phases with the corresponding ordering wave vector indicated. Thick black lines are the classical phase boundaries. The red area is the regime where Ref.~\cite{Ghosh-2019} identifies a non-magnetic phase. Red points mark the sets of Heisenberg couplings considered here. On the left and right sides of the phase diagram we depict the states with ordering wave vectors ${\mathbf q}=(2\pi,0,0)$ and ${\mathbf q}=(\pi,\pi,\pi)$. Note that in the ${\mathbf q}=(2\pi,0,0)$ state the two sublattices have opposite spin orientations. For the ${\mathbf q}=(\pi,\pi,\pi)$ order, the $B$ sublattice has the same spin configuration as the $A$ sublattice, but globally rotated by an angle $\pi/2$.}\label{fig:bcc_phase_diag}
\end{figure}

We now investigate longer-range mean-field terms in the BCC 1 case. As explained in Sec.~\ref{sr_example}, one can apply a certain gauge transformation such that on the nearest neighbor level the two groups of projective representations in Table~\ref{tab:psg_bcc_example} become indistinguishable. However, this is no longer possible for longer-range terms, i.e., when allowing for second and third neighbor amplitudes on top of the nearest neighbor model in Eq.~(\ref{eq:bcc_1a}) one needs to distinguish between these two cases. Particularly, for the PSGs in the first line, hopping and pairing amplitudes of second and third neighbor type may occur simultaneously:
\begin{align}\label{eq:bcc1_2a}
&\text{BCC 1$_2$:} \quad  u_{\delta \mathbf{r}}= \chi_{2}\tau^3 + \Delta_{2}\tau^1,\;\forall~\delta \mathbf{r} \text{ second neighbors}\;,\notag\\
&\text{BCC 1$_3$:} \quad u_{\delta \mathbf{r}}=\chi_{3}\tau^3 + \Delta_{3}\tau^1,\;\forall~\delta \mathbf{r} \text{ third neighbors}\;.
\end{align}
Here, second (third) neighbor bonds are of the form $\delta\mathbf{r}=(\pm1,0,0)$ ($\delta\mathbf{r}=(\pm1,\pm1,0)$), and permutations of coordinates. For the PSGs in the second line of Table~\ref{tab:psg_bcc_example}, the projective implementations of symmetries forbid spinon pairing terms, i.e., $\Delta_2=\Delta_3=0$. However, all our self-consistent calculations indicate that finite $\Delta_2$ and $\Delta_3$ terms significantly lower the mean-field energies $\epsilon$ such that the PSGs in the second line are energetically unfavorable. Therefore, we will not further discuss the case $\Delta_2=\Delta_3=0$ but focus on the more general type of ansatz in Eq.~(\ref{eq:bcc1_2a}). Note that the second neighbor terms in Eq.~(\ref{eq:bcc1_2a}) break the IGG down to $U(1)$ while the inclusion of third neighbor terms further reduces the IGG down to $\mathds{Z}_2$.

The terms in Eq.~(\ref{eq:bcc1_2a}) are self-consistently generated for spin models with frustrating antiferromagnetic second and third neighbor spin interactions $J_2$ and $J_3$. Here, we consider two special coupling scenarios for $J_2$ and $J_3$ where enhanced quantum fluctuations are expected, increasing the propensity for spin liquid behavior. The first case is given by $J_{2}/J_{1}=2/3$, $J_3=0$ where the corresponding classical spin system undergoes a phase transition between the aforementioned ${\mathbf q}=(2\pi,0,0)$ N\'eel state and a stripe ordered ${\mathbf q}=(\pi,\pi,\pi)$ phase~\cite{Utsumi-1977,Schmidt-2002,Farnell-2016,Jur-2020}, see the phase diagram in Fig.~\ref{fig:bcc_phase_diag}. In the second case, we consider $(J_{2}/J_{1},J_{3}/J_{1})=(2/3,1/4)$ where recent studies have identified a magnetically disordered phase~\cite{Ghosh-2019} (red area in Fig.~\ref{fig:bcc_phase_diag}).
\begin{figure}[b]
\includegraphics[width=\columnwidth]{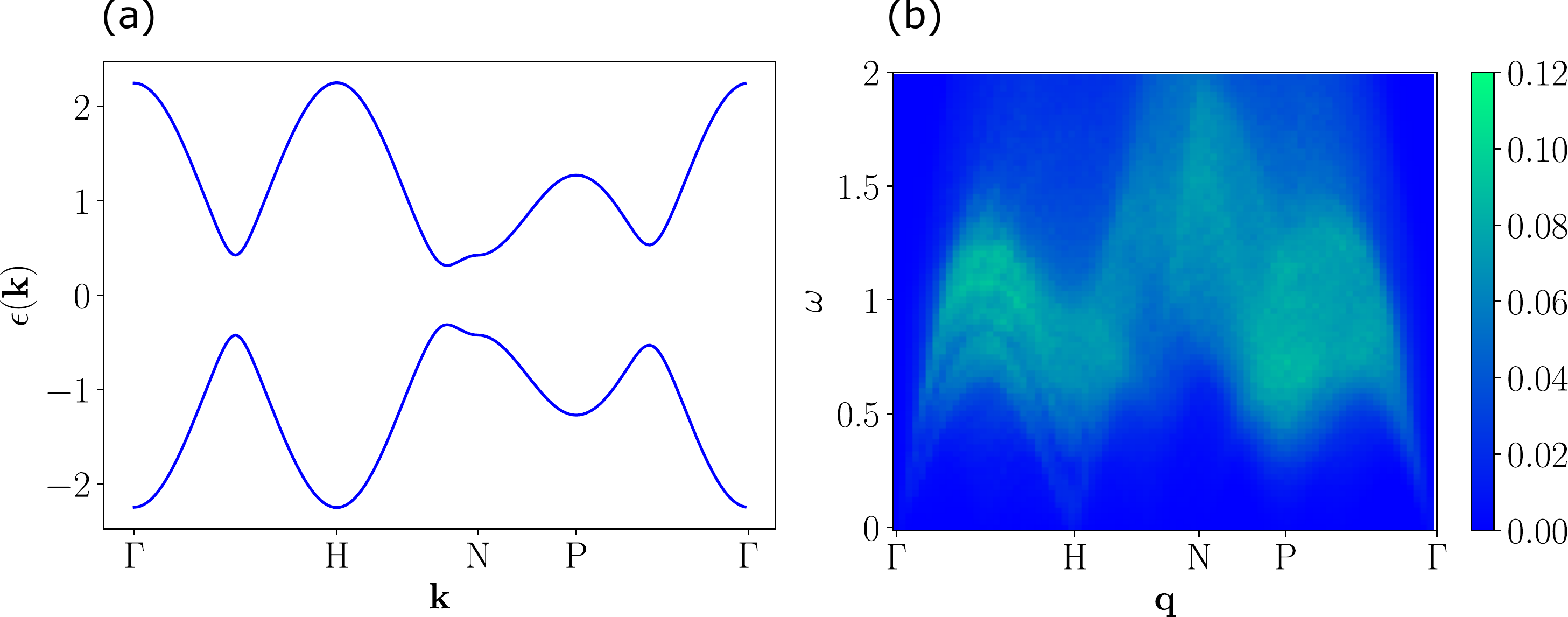}
\caption{``BCC 1$_2$'' model with mean-field amplitudes up to second neighbors. (a) Self-consistent spinon dispersion for the first neighbor terms in Eq.~\eqref{eq:bcc_1a} and second neighbor terms in Eq.~\eqref{eq:bcc1_2a} along a path through the first Brillouin zone. (b) Corresponding dynamical spin structure factor along the same path in reciprocal space.}\label{fig:bcc1_2a_state}
\end{figure}
\begin{figure*}
\includegraphics[width=\textwidth]{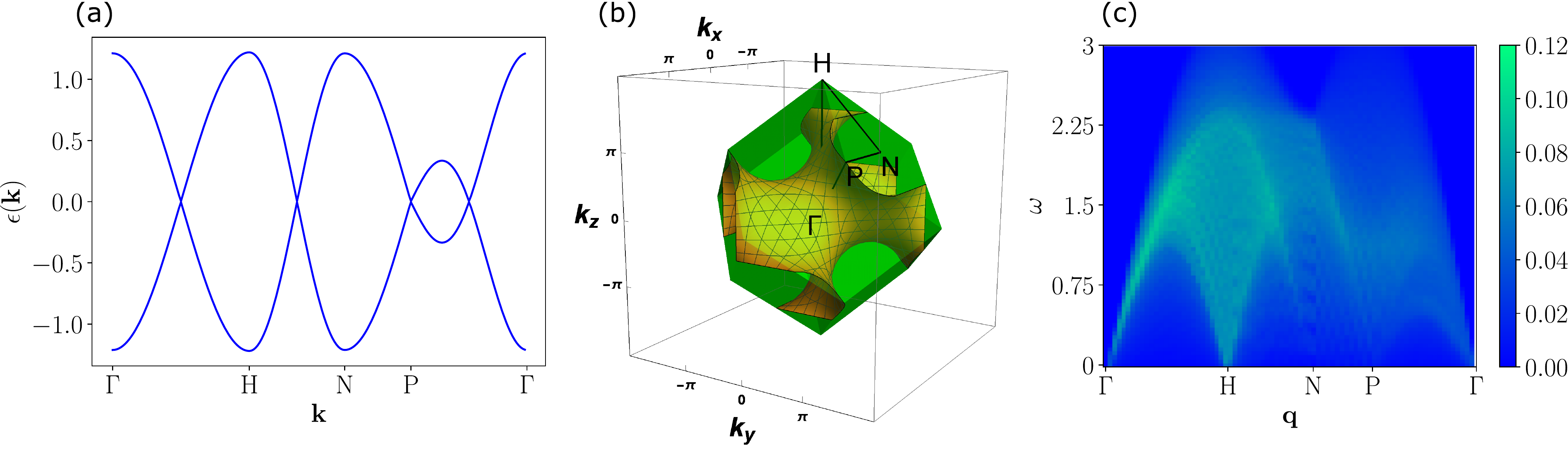}
\caption{Nearest neighbor model ``BCC 2$_1$''. (a) Spinon dispersion of Eq.~\eqref{eq:bcc2_1} along a path through the first Brillouin zone. The Fermi surface is depicted in (b) where the green region indicates the first Brillouin zone. (c) Dynamical spin structure factor along a path in reciprocal space.}\label{fig:bcc_2_state}
\end{figure*}

The self-consistently calculated amplitudes for $(J_{2}/J_{1},J_{3}/J_{1})=(2/3,0)$ are given by
\begin{align}
&\chi_1 =0.116\;,\;\chi_2 = -4.7\cdot10^{-4}\;,\; \Delta_2 = 0.106\;,\notag\\
&a_3 = -9.1\cdot10^{-4}\;,\;\epsilon=-0.178\;.
\end{align} 
The pairing term opens a gap in the spectrum as illustrated in Fig.~\ref{fig:bcc1_2a_state}(a). As a result, the cone like-signal around the $\Gamma$ point and the high intensities at the $H$ point disappear in the dynamical spin structure factor [see \ref{fig:bcc1_2a_state}(b)]. Instead a characteristic pattern of three arcs appears which are located along the lines $\overline{\Gamma H}$, $\overline{H P}$, and $\overline{P \Gamma}$.

In the second case where $(J_{2}/J_{1},J_{3}/J_{1})=(2/3,1/4)$ we find small additional $\chi_3$ and $\Delta_3$ terms while the other amplitudes remain nearly unchanged:
\begin{align}
&\chi_1 =0.116\;,\;\chi_2 = -2.8\cdot10^{-4}\;,\; \Delta_2 = 0.105\;,\notag\\
&\chi_3=-8.7\cdot10^{-5}\;,\;\Delta_3=-0.014\;,\;a_3 = -9.1\cdot10^{-4}\;,\notag\\
&\epsilon=-0.178\;.
\end{align}
As compared to Fig.~\ref{fig:bcc1_2a_state} these modifications only marginally modify the spinon spectrum and the spin structure factor, indicating that this spin liquid phase is rather insensitive with respect to $J_3$ interactions. Hence, the spin structure factor in Fig.~\ref{fig:bcc1_2a_state}(b) represents the characteristic magnetic response in the BCC 1 case.

\subsubsection{BCC 2: $\eta_P = -1$ state}
In the case $\eta_P = -1$, the nearest neighbor hopping amplitudes have a direction-dependent sign structure induced by a non-trivial action of the transformations $\Pi_z$, $\Pi_y$ and $P$:
\begin{align}\label{eq:bcc2_1}
\text{BCC 2$_1$:} \quad &u_{(1/2, 1/2, 1/2 )}=\chi_{1}\tau^3 = u_{(-1/2, -1/2, -1/2 )} \notag \\
& = u_{(1/2, -1/2, 1/2 )} = u_{(-1/2, 1/2, -1/2 )}\notag \\
& = u_{(1/2, 1/2, -1/2 )} = u_{(-1/2, -1/2, 1/2 )}\notag \\
& = -u_{(-1/2, 1/2, 1/2 )} = -u_{(1/2, -1/2, -1/2 )}\;,\notag \\
& a_3 \neq  0\;.
\end{align}
As can be seen, one of the four nearest neighbor directions carries hopping amplitudes with opposite signs. The IGG of this ansatz remains $SU(2)$. The self-consistent mean-field parameters and energy per site for a nearest neighbor coupling $J_1=1$ are given by
\begin{equation}
\chi_1=0.152\;,\quad a_3=-0.0045\;,\quad \epsilon =-0.208\;.
\end{equation}
Most importantly, already on the nearest neighbor level, the energy of this state is significantly lower than for the BCC 1 case. The corresponding spinon dispersion shown in Fig.~\ref{fig:bcc_2_state}(a) features a Fermi surface which has an entirely different shape compared to the nearest neighbor BCC 1 ansatz. This also reflects in the dynamical spin structure factor which, in absence of any nesting vectors, exhibits a more evenly distributed intensity with a characteristic arc emanating from the $\Gamma$-point and reaching its maximum at the $H$-point [Fig.~\ref{fig:bcc_2_state}(c)]. In contrast to the BCC 1 case much of the total weight appears between $\Gamma$ and $H$ while the region between $\Gamma$ and $P$ shows a relatively small signal. 

The projective implementation of symmetries in this PSG, characterized by the sign factors $\eta_\Pi = \eta_P = -1$, dictates that no second neighbor mean-field terms are allowed. This also implies that when adding second neighbor $J_2$ interactions, the results from the $J_1$-only case remain unchanged. Third neighbor terms can exist and similarly to the BCC 1 case one needs to distinguish between the two representations in Table~\ref{tab:psg_bcc_example}. For the PSG in the first line, the third neighbor terms include spinon hopping and pairing of the form
\begin{align}\label{eq:bcc2_3a}
\text{BCC 2$_3$:} \quad &u_{(1, 1, 0)} = \chi_3 \tau^3 + \Delta_3 \tau^1 = u_{(-1, -1, 0)} \notag \\
& = u_{(0, 1, 1)} = u_{(0, -1, -1)} = u_{(1, 0, -1)} = u_{(-1, 0, 1)} \notag \\
& = -u_{(1, 0, 1)} = -u_{(-1, 0, -1)} = -u_{(1, -1, 0)} \notag \\
& = -u_{(-1, 1, 0)} = -u_{(0, 1, -1)} = -u_{(0, -1, 1)}\;,
\end{align}
while for the PSG in the second line the pairing terms are forbidden, $\Delta_3=0$. Since we again find that a finite $\Delta_3$ lowers the energy compared to $\Delta_3=0$ we only treat the more general case where spinon hoppings and pairings are both present. Note that similar to the first neighbor amplitudes in Eq.~(\ref{eq:bcc2_1}) the third neighbor terms show a direction dependent sign pattern.

\begin{figure}
\includegraphics[width=\columnwidth]{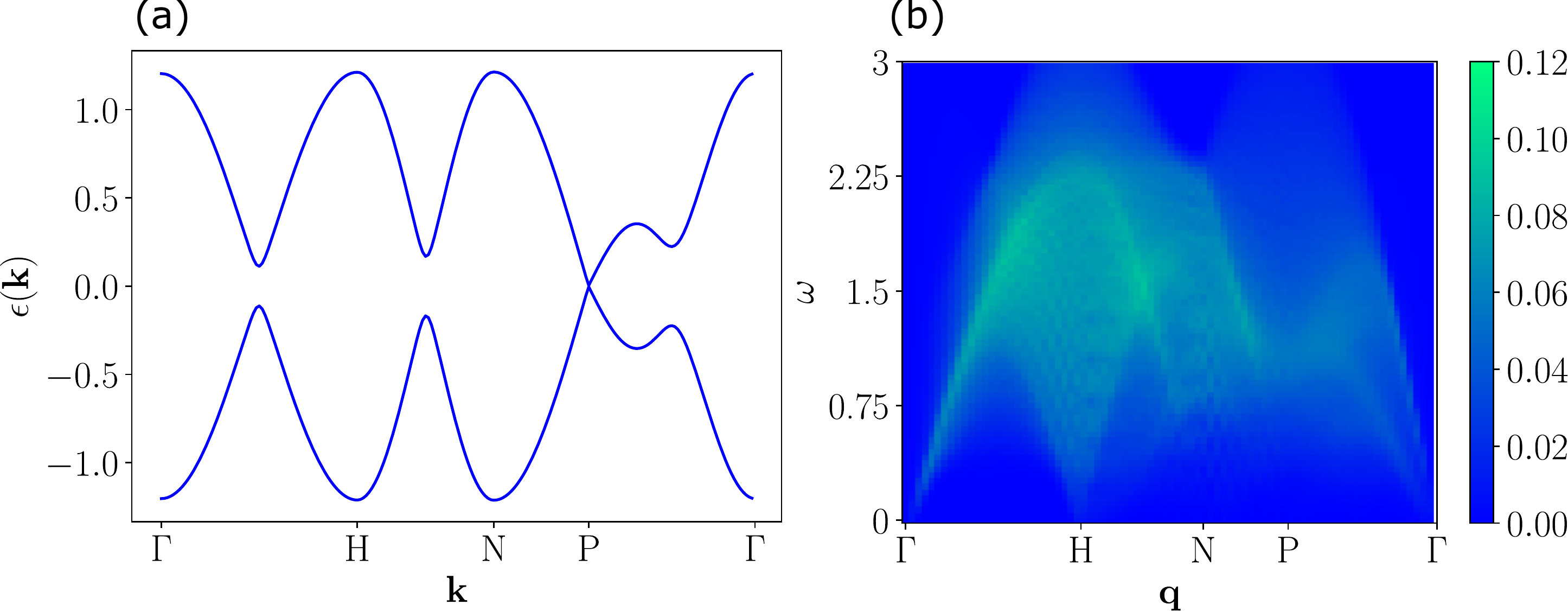}
\caption{``BCC 2$_3$'' model with mean-field amplitudes up to third neighbors. (a) Self-consistent spinon dispersion for the first neighbor terms in Eq.~\eqref{eq:bcc2_1} and third neighbor terms in Eq.~\eqref{eq:bcc2_3a} along a path through the first Brillouin zone. (b) Corresponding dynamical spin structure factor along the same path in reciprocal space.}\label{fig:bcc2_3a_state}
\end{figure}

The self-consistent mean-field amplitudes for $(J_{2}/J_{1},J_{3}/J_{1})=(2/3,1/4)$ given by
\begin{align}
&\chi_1 =0.151\;,\;\chi_3=-5.8\cdot10^{-4}\;,\; \Delta_3=0.0283\;,\notag\\
&a_3 = -0.0037\;,\;\epsilon=-0.209
\end{align}
differ only slightly from the $J_1$-only case, however, the finite $\Delta_3$ term breaks the IGG down to $\mathds{Z}_2$. The pairing term gaps out parts of the Fermi surface but leaves behind a nodal Dirac point at $P=(\pi, \pi, \pi)$ [Fig.~\ref{fig:bcc2_3a_state}(a)]. Due to the smallness of $\Delta_3$, the dynamical spin structure factor, shown in Fig.~\ref{fig:bcc2_3a_state}(b), deviates from the one in Fig.~\ref{fig:bcc_2_state}(c) only at low energies where the signal is suppressed. Since the mean-field energies are significantly smaller compared to the BCC 1 case, this analysis suggests that Figs.~\ref{fig:bcc_2_state}(c) and \ref{fig:bcc2_3a_state}(b) represent typical intensity distributions of the spin structure factor for possible quantum spin liquids on the bcc lattice. 

An overview of the short-range mean field models and the corresponding projective implementations of symmetries can be found in Table~\ref{tab:bcc_summary}.
\subsection{Face centered cubic lattice}
\begin{figure*}
\includegraphics[width=\textwidth]{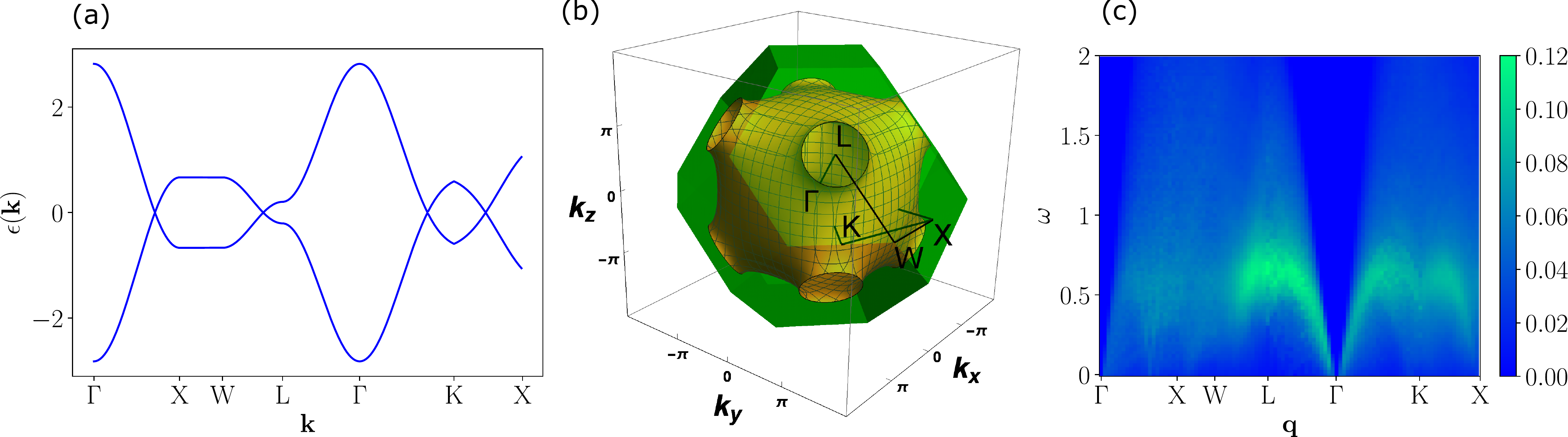}
\caption{Nearest neighbor model ``FCC 1$_1$''. (a) Spinon dispersion of Eq.~\eqref{eq:fcc_1_1} along a path through the first Brillouin zone where $X=(0,2\pi,0)$, $W=(\pi,2\pi,0)$, $L=(\pi,\pi,\pi)$, and $K=(\frac{3}{2}\pi,\frac{3}{2}\pi,0)$ (and symmetry related wave vectors). The Fermi surface is depicted in (b) where the green region indicates the first Brillouin zone. (c) Dynamical spin structure factor along a path in reciprocal space.}\label{fig:fcc_1_state}
\end{figure*}
We finally treat the fcc lattice where a classification of PSGs on the nearest neighbor level leads to four different cases listed in Table~\ref{tab:psg_fcc}. Similar to the bcc lattice in the previous section one can perform a gauge transformation generated by $\tau^2$ such that the nearest neighbor ans\"atze in the first two lines become identical (this, however, does not work for longer-range amplitudes). Furthermore, the third and fourth lines yield ans\"atze which can be transformed into each other by a simple permutation of the cartesian axes. Consequently, only two nearest neighbor cases need to be considered, where the projective action of $P$ is implemented as $g_P=\tau^0$ or as $g_P=e^{i\frac{\pi}{3}\tau^2}$.
\begin{table}[h!]
\setlength{\tabcolsep}{0.3cm}
\begin{tabular}{ >{$}c<{$}  >{$}c<{$}  >{$}c<{$}  >{$}c<{$} }
\hline\hline
\eta_{\mathcal{T}}g_\mathcal{T} & \eta_P g_P & \eta_I g_I & \eta_{\Pi_{xy}} g_{\Pi_{xy}} \\
\hline
+i\tau^2  & +\tau^0 & +\tau^0 & +\tau^0  \\
+i\tau^2  & +\tau^0 & +\tau^0/+i\tau^3 & +\tau^0/+i\tau^3 \\
+i\tau^2  & +e^{i\frac{\pi}{3}\tau^2} & +\tau^0 & +i\tau^3 \\
+i\tau^2  & +e^{i\frac{2\pi}{3}\tau^2} & +\tau^0 & +i\tau^3 \\
\hline\hline
\end{tabular}
\caption{Possible PSG representations for first neighbor ans\"atze on the fcc lattice. Note that in the second line at least one of the matrices $g_I$, $g_{\Pi_{xy}}$ must be given by $i\tau^3$.}\label{tab:psg_fcc}
\end{table}

\subsubsection{FCC 1: $g_P = \tau^0$ state}
\begin{figure}[b]
\includegraphics[width=0.9\linewidth]{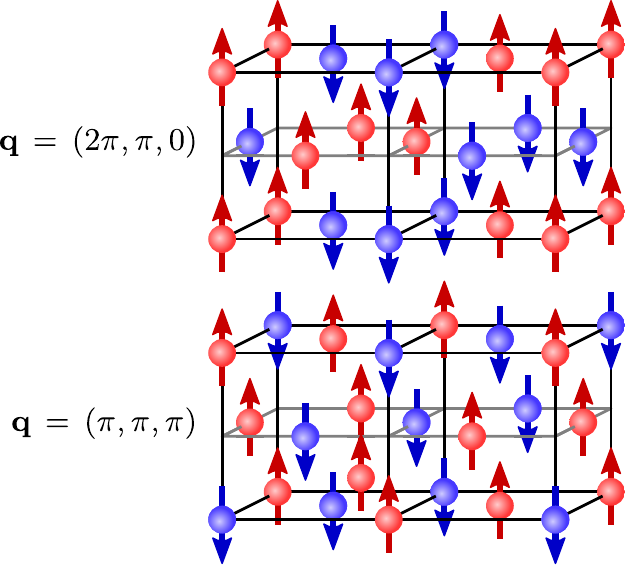}
\caption{Relevant magnetic orders of an antiferromagnetic classical $J_1$-$J_2$ Heisenberg model on the fcc lattice: At $J_2/J_1=0.5$ the ${\mathbf q}=(2\pi,\pi,0)$ magnetic order (top) shows a phase transition into ${\mathbf q}=(\pi,\pi,\pi)$ magnetic order (bottom).}\label{fig:fcc_lattice}
\end{figure}
We again start our discussion with first neighbor ans\"atze and then add terms up to third neighbors. A Heisenberg model on the fcc lattice with only nearest neighbor spin interactions $J_1$ is already frustrated and there are, indeed, numerical studies predicting a magnetically disordered state~\cite{Kuzmin-2003,revelli19}. The enhanced quantum fluctuations in this model stem from the fact that the corresponding classical spin system exhibits lines in reciprocal space along which the ground state energies are degenerate~\cite{Smart-1966}.

The ansatz class with $g_P = \tau^0$, represented by the first and second lines of Table~\ref{tab:psg_fcc}, consists of a uniform hopping on nearest neighbor bonds,
\begin{align}\label{eq:fcc_1_1}
\text{FCC 1$_1$}: \quad &u_{\delta \mathbf{r}}=\chi_{1}\tau^3, \quad \forall~\delta \mathbf{r} \text{ first neighbors}\;, \notag \\
&a_3 \neq  0\;,
\end{align}
where $\delta\mathbf{r}=(\pm1/2,\pm1/2,0)$ (and permutations of coordinates) and the IGG is $U(1)$. We find the following self-consistent mean-field amplitudes and ground state energy for $J_{1}=1$:
\begin{equation}
\chi_1 = 0.109\;,\quad a_3 =0.204\;,\quad \epsilon= -0.156\;.
\end{equation}
This ansatz has a spinon dispersion and Fermi surface shown in Figs.~\ref{fig:fcc_1_state}(a) and~\ref{fig:fcc_1_state}(b). The dynamical spin structure factor in Fig.~\ref{fig:fcc_1_state}(c) exhibits a rather homogeneous distribution of magnetic response where the flanks of a cone around the $\Gamma$ point form a region of larger signal.

When adding second and third neighbor mean-field amplitudes one needs to distinguish between the first two lines of Table~\ref{tab:psg_fcc}. Similar to the BCC 1 case, the first line allows for a more general ansatz with spinon hopping and pairing
\begin{align}\label{eq:fcc_1_2a}
&\text{FCC 1$_2$}: \quad  u_{\delta \mathbf{r}}=\chi_{2}\tau^3+\Delta_2\tau^1,\;\forall~\delta \mathbf{r} \text{ second neighbors}\;,\notag\\
&\text{FCC 1$_3$}: \quad u_{\delta \mathbf{r}}=\chi_{3}\tau^3+\Delta_3\tau^1,\;\forall~\delta \mathbf{r} \text{ third neighbors}\;,
\end{align}
while for the second line one finds $\Delta_2=\Delta_3=0$. Here, second (third) neighbor bonds are of the form $\delta\mathbf{r}=(\pm1,0,0)$ ($\delta\mathbf{r}=(\pm1/2,\pm1/2,\pm1)$), and permutations of coordinates. Due to the same reason as for the bcc lattice, we treat $\Delta_2$ and $\Delta_3$ as being finite, in which case the IGG is broken down to $\mathds{Z}_2$.

We consider two sets of longer-range spin interactions: A first interesting physical scenario appears when $(J_{2}/J_{1},J_{3}/J_{1})=(0.5,0)$. As a function of $J_2/J_1$ this point marks the phase transition in the corresponding classical model between magnetic phases with ordering vectors $W=(2\pi,\pi,0)$ and $L=(\pi,\pi,\pi)$, see Fig.~\ref{fig:fcc_lattice} for an illustration of these orders. Interestingly, the manifold of degenerate ground states for these couplings is even enlarged compared to the $J_1$-only case, forming surfaces in momentum space~\cite{Balla-2019} The second set of couplings is given by $(J_2/J_1,J_3/J_1)=(0.5,0.25)$ where the classical model exhibits a triple point between magnetic phases with commensurate ordering vectors $X=(2\pi,0,0)$ and $L=(\pi,\pi,\pi)$ as well as an incommensurate spiral with ${\mathbf q}=(q,0,0)$~\cite{prep}. Hence, both sets of couplings promote quantum fluctuations and appear very promising for finding quantum spin liquid phases~\cite{Ignatenko-2008,revelli19}. 

\begin{figure}[b]
\includegraphics[width=\columnwidth]{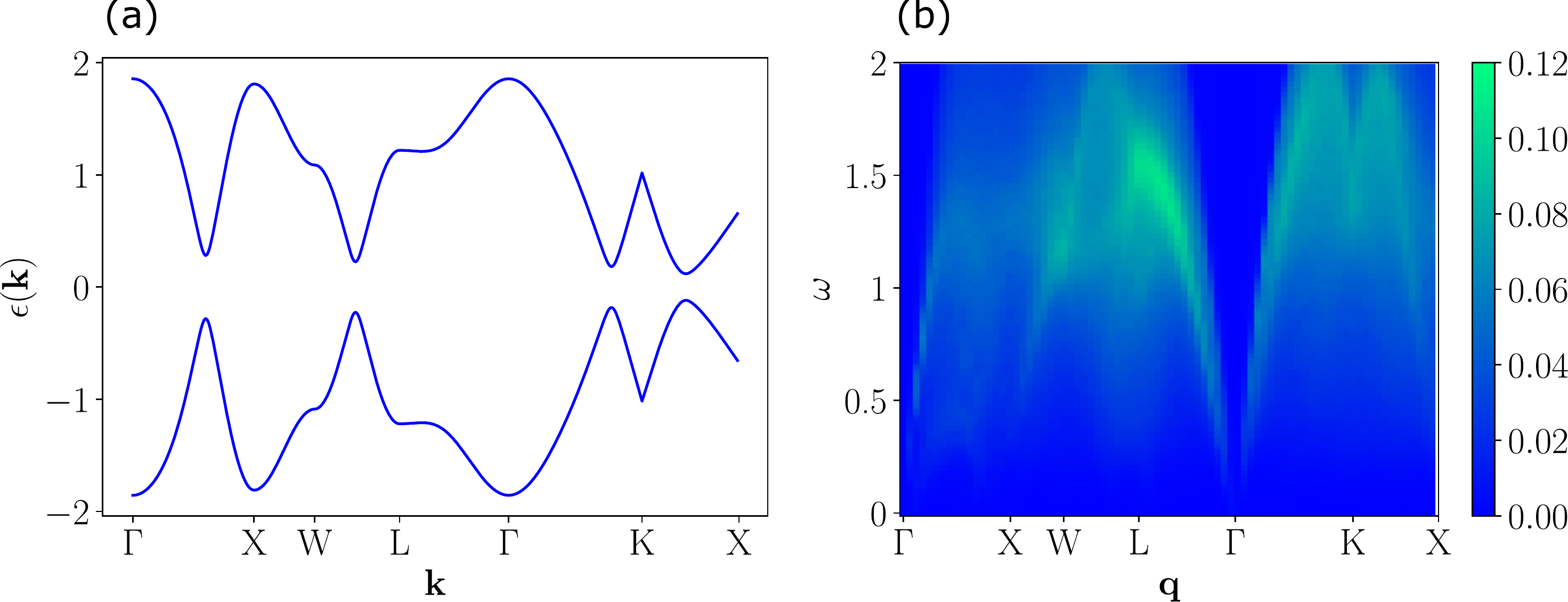}
\caption{``FCC 1$_2$'' model with mean-field amplitudes up to second neighbors. (a) Self-consistent spinon dispersion for the first neighbor terms in Eq.~\eqref{eq:fcc_1_1} and second neighbor terms in Eq.~\eqref{eq:fcc_1_2a} along a path through the first Brillouin zone. (b) Corresponding dynamical spin structure factor along the same path in reciprocal space.}\label{fig:fcc_2a_state}
\end{figure}

Solving the self-consistent equations for $(J_2/J_1,J_3/J_1)=(0.5,0)$ yields the amplitudes and energy
\begin{align}
&\chi_1 =0.106\;,\;\chi_2 = -0.075\;,\; \Delta_2 = 0.059\;,\notag\\
&a_3 = 0.09\;,\;\epsilon=-0.185\;.
\end{align}
The additional $\Delta_2$ term gaps out the spinon dispersion, see Fig.~\ref{fig:fcc_2a_state}(a). Since $\chi_2$ and $\Delta_2$ are both non-negligible they also have a significant effect on the spinon dispersion away from the points of gap opening. As a result of the spinon gap, the V-shaped signal in the dynamical spin structure factor at the $\Gamma$ point becomes less pronounced but still represents the most salient feature [see Fig.~\ref{fig:fcc_2a_state}(b)]. The second set of Heisenberg interactions $(J_2/J_1,J_3/J_1)=(0.5,0.25)$ yields somewhat modified mean-field amplitudes with a slightly lower energy
\begin{align}
&\chi_1 =0.106\;,\;\chi_2 = -0.066\;,\; \Delta_2 = 0.067\;,\notag\\
&\chi_3=-0.028\;,\;\Delta_3=-0.0132\;,\;a_3 = 0.093\;,\notag\\
&\epsilon=-0.192\;.
\end{align}
The corresponding spinon dispersion and dynamical spin structure factor, however, are qualitatively similar to the previous model.
\subsubsection{FCC 2: $g_P = e^{i\frac{\pi}{3}\tau^2}$ state}
\begin{figure*}
\includegraphics[width=\textwidth]{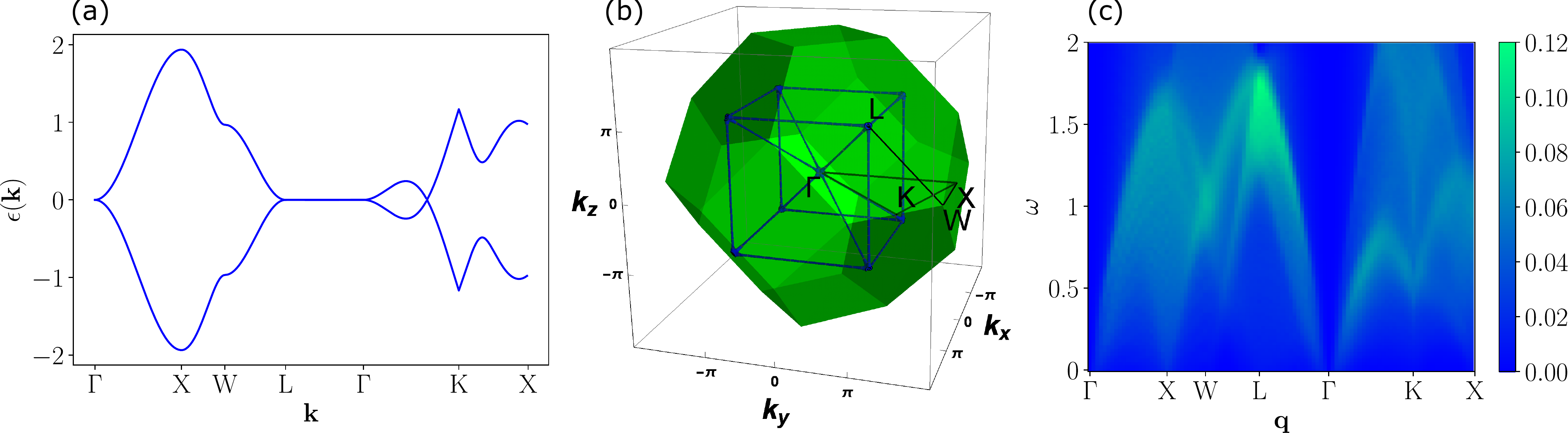}
\caption{Nearest neighbor model ``FCC 2$_1$''. (a) Spinon dispersion of Eq.~\eqref{eq:fcc_2_1} along a path through the first Brillouin zone. Note the symmetry protected zero energy modes along the line $\overline{\Gamma L}$. (b) Fermi lines emanating from the $\Gamma$-point and forming a cube-like pattern. The green region indicates the first Brillouin zone. (c) Dynamical spin structure factor along a path in reciprocal space.}\label{fig:fcc_2_state}
\end{figure*}
The second type of ans\"atze on the fcc lattice has a richer structure, characterized by the nearest neighbor terms
\begin{align}\label{eq:fcc_2_1}
\text{FCC 2$_1$}: \quad& u_{(\pm1/2,\pm1/2,0)} = \chi_{1} \tau^3\;, \notag \\
& u_{(\pm1/2,0,\pm1/2)} = \chi_{1} \left( \frac{\sqrt{3}}{2} \tau^1 -\frac{1}{2}\tau^3 \right)\;, \notag \\
& u_{(0,\pm1/2,\pm1/2)} = \chi_{1} \left( -\frac{\sqrt{3}}{2} \tau^1 -\frac{1}{2}\tau^3 \right)\;,
\end{align}
where $(\pm1/2,\pm1/2,0)$ denotes the four bonds $(1/2,1/2,0)$, $(1/2,-1/2,0)$, $(-1/2,1/2,0)$, $(-1/2,-1/2,0)$ and equivalently for the other terms. The non-trivial matrix structure of $g_P$ induces an interesting connection between real space and spinor space transformations: While the nearest neighbor bonds in the three lines of Eq.~(\ref{eq:fcc_2_1}) are related by $2\pi/3$-rotations around the $(1,1,1)$ axis, the terms on the right hand sides transform into each other under $2\pi/3$-rotations around the $\tau^2$ axis in the space of mean-field matrices. Here, we have chosen a gauge in which the $(\pm1/2,\pm1/2,0)$ bonds only carry hopping amplitudes. Due to the special projective action of $P$, the other bonds then carry a combination of hopping and pairing such that even on the nearest neighbor level the gauge structure is $\mathds{Z}_2$. 

The projective action of $P$ has consequences on the spinon dispersion, independent of the range of mean-field amplitudes. For momenta $\mathbf{k}^*\equiv(k^*,k^*,k^*)=P(\mathbf{k}^*)$ which map back onto itself under permutation, the mean-field Bloch Hamiltonian $H_\text{mf}(\mathbf{k})$ needs to fulfill the relation $g_P^{-1} H_\text{mf}(\mathbf{k}^*) g_P = H(\mathbf{k}^*)$. On the other hand, the combination of time reversal $\mathcal{T}$ and inversion $I$ leads to an additional condition $(g_Ig_{\mathcal{T}})^{-1}H_{\text{mf}}^*(\mathbf{k})g_Ig_{\mathcal{T}}=-H_{\text{mf}}(\mathbf{k})$ where it has been used that $I\mathcal{T}$ leaves any momentum $\mathbf{k}$ invariant. This means that for momenta $\mathbf{k}^*=(k^*,k^*,k^*)$ and real Bloch Hamiltonians (as considered here), $H_{\text{mf}}(\mathbf{k}^*)$ has to commute with $g_P = e^{i\frac{\pi}{3}\tau^2}$ but anti-commute with $g_Ig_{\mathcal{T}}=i\tau^2$. Since this can only be fulfilled for $H_{\text{mf}}(\mathbf{k}^*)=0$ the system features zero-energy modes along the line $\overline{\Gamma L}$ running through the entire Brillouin zone. Similar arguments can be formulated for all momenta $\mathbf{k}=(\pm k,\pm k,\pm k)$ such that the spinon dispersion shows a symmetry protected star-like pattern of zero energy lines emanating from the $\Gamma$ point. Additionally, by analyzing the actions of the sublattice translations $t_1$ and $t_2$ one obtains a condition according to which another network of line-like zero modes forming a cube structure with corners at the $L$-points exists. This is illustrated in Figs.~\ref{fig:fcc_2_state}(a) and \ref{fig:fcc_2_state}(b) where a nearest neighbor Heisenberg model with $J_1=1$ is considered leading to the self-consistent mean-field amplitude $\chi_1 = 0.121$. Particularly, Fig.~\ref{fig:fcc_2_state}(b) shows the star and cube-like pattern of modes at the Fermi level. Note that no symmetry allowed Lagrange multipliers are possible. We find a mean-field energy per site of $\epsilon = -0.198$ which is significantly lower compared to the FCC 1 case.

The cube-like network of zero-modes can be mapped onto itself by nesting vectors of the type $X=(2\pi,0,0)$. As a consequence, the dynamical spin structure factor in Fig. \ref{fig:fcc_2_state}(c) shows a faint signal of low energy response at the $X$-point. The weakness of this feature compared to the strong nesting signal of the BCC 1$_1$ ansatz in Fig.~\ref{fig:bcc_1_state} can be explained by the fact that here, the nesting occurs along lines and not along planes. As an additional characteristic feature of the FCC 2 case, the dynamical spin structure factor shows a spot of high intensity at the $L$-point, marking the upper edge of the excitation spectrum.

No second neighbor terms can be included without violating the projective symmetries.

Finally, the third neighbor terms follow a similar scheme as the nearest neighbor ones:
\begin{align}\label{eq:fcc_2_3}
\text{FCC 2$_3$}: \quad& u_{(\pm1/2,\pm1/2,\pm1)} = \chi_3 \tau^3\;, \notag \\
& u_{(\pm1/2,\pm1,\pm1/2)} = \chi_3 \left( \frac{\sqrt{3}}{2} \tau^1 -\frac{1}{2}\tau^3 \right)\;, \notag \\
& u_{(\pm1,\pm1/2,\pm1/2)} = \chi_3 \left( -\frac{\sqrt{3}}{2} \tau^1 -\frac{1}{2}\tau^3 \right)\;.
\end{align}
Considering again the spin interactions $(J_2/J_1,J_3/J_1)=(0.5,0.25)$ we find the following mean-field parameters and energy per site:
\begin{equation}
\chi_1 =  0.121\;,\quad\chi_3 =-0.035\;,\quad\epsilon=-0.208\;.
\end{equation}
While the spinon dispersion and dynamical spin structure factor are similar to the nearest neighbor ansatz (with the zero modes preserved) it is worth highlighting that the energy is again smaller than in the FCC 1 case, indicating that at least on the mean-field level this spin liquid phase appears energetically preferred.

An overview of the short-range mean field models and the corresponding projective implementations of symmetries for the fcc lattice can be found in Table~\ref{tab:fcc_summary}.

\section{Discussion and conclusion}\label{conclusion}
The three lattices considered in this work are characterized by large numbers of elements of their symmetry groups. Therefore, it is not surprising that our PSG classifications of spin liquid phases yield a plethora of possible projective representations which even exceeds a thousand for the bcc lattice. However, the large numbers of symmetries also imply that short-range mean-field ans\"atze are subject to many constraints and, as a consequence, only two possible nearest neighbor models remain for each of the three lattices. Even though the exact amount of PSGs depends on the precise group algebra, we conclude that the systems considered here feature a particularly marked discrepancy between the number of algebraic PSGs and the number of short range mean-field ans\"atze. As an example, one may compare this with the 2D kagome lattice where the symmetry group has only four generators (two translations and two point group operations). There, one finds 20 PSGs which reduce to four nearest neighbor ans\"atze~\cite{Lu-2011,Hering-2019}.

The two nearest neighbor models which we identify for each of the three lattices share the common property that one of them exhibits simple uniform spinon hopping while the other features hopping amplitudes with special sign patterns or a particular locking between spinon hopping and pairing (see the FCC 2 state). These spatial modulations are caused by non-trivial projective actions of translations $T_\mu$ or permutation $P$. Interestingly, already on the nearest neighbor level these non-trivial ans\"atze are the ones with the lowest mean-field ground state energy and the addition of longer-range amplitudes does not qualitatively change this behavior. We further discuss characteristic features in the spin structure factor which allow one to distinguish these states.

One overall assumption of our study is that the mean-field amplitudes are always time reversal and spin-rotation invariant. When starting from a Heisenberg Hamiltonian as in Eq.~(\ref{ham}) this seems justified, however, it can generally not be excluded that these symmetries are broken {\it spontaneously} even in quantum spin liquids, which leads to so-called chiral~\cite{Becca-2015} or nematic~\cite{Lu-2016nem} spin liquids, respectively. The scenario of chiral spin liquids appears unusual in our systems as they preferably form when quantum fluctuations melt non-coplanar classical spin orders~\cite{Messio-2011,Bieri-2016}. For Heisenberg models on Bravais lattices as considered here, however, all classical ground states are coplanar or even collinear. Similarly, while nematic spin liquid ground states are unusual in spin-1/2 systems with only antiferromagnetic Heisenberg interactions~\cite{Iqbal-2019prx}, recent studies in 2D indicate that frustrating antiferromagnetic {\it and} ferromagnetic couplings may induce such a scenario~\cite{Shannon-2006,Momoi-2006,Iqbal-2016}. Additionally, a multitude of further spin liquid phases may be constructed when assuming that spin-rotation invariance is already broken on the level of the spin Hamiltonian, e.g. through Dzyaloshinskii-Moriya interactions (due to the systems' inversion symmetries such terms would, however, not be allowed on nearest neighbor bonds). We leave such extensions for future studies.

Also from a methodological perspective it is clear that our work rather represents a first step towards more refined studies. For example, our ground state energies and dynamical spin structure factors are certainly subject to a mean-field bias and the gauge fluctuations which we neglect may lead to a smearing of otherwise sharp features in the magnetic response~\cite{Punk-2014}. (We note, however, that for the $\mathds{Z}_2$ gauge structures considered here, the mean-field biases are expected to be smaller compared to $U(1)$ or $SU(2)$ scenarios). The limitations of mean-field can be overcome when using our PSGs as an input for variational Monte Carlo. By Gutzwiller-projecting fermionic parton wave functions this technique allows one to faithfully calculate ground state energies well beyond mean-field~\cite{Iqbal-2013,Iqbal-2016tri}. Likewise, the Gutzwiller projections enable the calculation of more accurate dynamical spin structure factors~\cite{Ferrari-2018a,Ferrari-2018b,Ferrari-2019,Ferrari-2020a,Mei-2015} which amounts to taking into account time-like fluctuations in the gauge fields. An alternative extension of our work is the combination with a functional renormalization group treatment as has recently been demonstrated in Ref.~\cite{Hering-2019}. In this scheme the spinon hopping and pairing amplitudes are subject to a renormalization group flow which takes into account dressed vertex functions instead of the bare interactions $J_{\mathbf{rr'}}$ considered here. Each of these extensions promise a more accurate and detailed investigation of quantum spin liquids in three dimensions.

\section*{Acknowledgements}
J.S. gratefully acknowledges the hospitality of the Indian Institute of Technology Madras, Chennai, India, where this project was initiated and parts of the work performed. J.S. further thanks Christian Kl\"ockner and Max Geier for discussions and useful comments. Y.I. thanks F. Becca, H. O. Jeschke, T. M\"uller, J. Richter, R. Thomale, and M. Zhitomirsky for helpful discussions and collaboration on related topics. This work was partially supported by the German Research Foundation within the CRC183 (project A02). Y.I. acknowledges the Science and Engineering Research Board (SERB), DST, India for support through Startup Research Grant No. SRG/2019/000056 and MATRICS Grant No. MTR/2019/001042, and the Abdus Salam International Centre for Theoretical Physics (ICTP) for support through the Simons Associateship scheme. This research was supported in part by the International Centre for Theoretical Sciences (ICTS) during a visit for participating in the program - Novel phases of quantum matter (Code: ICTS/topmatter2019/12) and in the program - “The 2nd Asia Pacific Workshop on Quantum Magnetism” (Code: ICTS/apfm2018/11). This research was supported in part by the National Science Foundation under Grant No. NSF PHY-1748958.

\appendix

\section{Projective implementation of lattice translations}\label{ap:gauge}
Here, we derive the gauge transformations associated with translations $T_x$, $T_y$, $T_z$. To realize the special gauge used in this work where $G_{T_\mu}$ [see Eq.~\eqref{eq:u_gauge}] is proportional to the identity matrix, we start with a local gauge transformation which acts on an ansatz according to $u_{\mathbf{rr'}} \rightarrow \tilde{u}_{\mathbf{rr'}}=W^\dagger_\mathbf{r} u_{\mathbf{rr'}} W_\mathbf{r'}$ (the new gauge is indicated by a tilde). Using Eq.~\eqref{eq:u_gauge} and inserting the identity $W_{\mathcal{S}(\mathbf{r})}W^\dagger_{\mathcal{S}(\mathbf{r})}$ twice one sees that a gauge transformation $W_\mathbf{r}$ also changes the projective implementations $G_{\mathcal{S}}$ of symmetry operations $\mathcal{S}$:
\begin{align*}
G_{\mathcal{S}}(\mathbf{r})\rightarrow\tilde{G}_{\mathcal{S}}(\mathbf{r}) = W^{\dagger}_{\mathbf{r}}G_{\mathcal{S}}(\mathbf{r})W_{S^{-1}(\mathbf{r})}\;.
\end{align*}
Starting at a given reference site $\mathbf{r}_0 = (x_0, y_0, z_0)$ one can use this local gauge freedom to enforce $\tilde{G}_{T_x}(\mathbf{r}_x)=\tau^0$ along the line $\mathbf{r}_x =(x, y_0, z_0)$. In the first step one finds
\begin{align*}
\tilde{G}_{T_x}(\mathbf{r}_0) = W^{\dagger}_{\mathbf{r}_0}G_{T_x}(\mathbf{r}_0)W_{\mathbf{r}_0 - \hat{e}_x}\overset{!}{=}\tau^0\;,
\end{align*}
where $\hat{e}_x$ denotes the unit vector in $x$-direction. This fixes $W_{\mathbf{r}_0 - \hat{e}_x} = G^{-1}_{T_x}(\mathbf{r}_0)W_{\mathbf{r}_0}$ and by successive applications of gauge transformations one finds $W_{\mathbf{r}_0 -n\hat{e}_x}=G^{-1}_{T_x}(\mathbf{r}_0-(n-1)\hat{e}_x)\ldots G^{-1}_{T_x}(\mathbf{r}_0) W_{\mathbf{r}_0}$ for the entire line. The same procedure can be performed for gauge transformations associated with $T_y$ for lines along the $y$-direction starting from any point on the line $\mathbf{r}_x$. This fixes the gauge $\tilde{G}_{T_y}(\mathbf{r}_{xy})=\tau^0$ in the plane $\mathbf{r}_{xy} = (x, y, z_0)$. Finally, one can enforce $\tilde{G}_{T_z}(\mathbf{r})=\tau^0$ on the entire lattice by starting at any point of the plane $\mathbf{r}_{xy}$. The local gauge freedom has thus been reduced to a global one given by the freedom to choose $W_{\mathbf{r}_0}$.

We continue in this gauge and determine all projective representations $G_{T_\mu}$ which are not yet fixed. To simplify the notation we omit the tilde in the following. Consider the sequence of translations $T_y T_z T^{-1}_y T^{-1}_z =id$ which requires a projective representation such that $G_{T_y}T_y G_{T_z} T_z T_y^{-1}G_{T_y}^{-1} T_z^{-1} G_{T_z}^{-1} \in \text{IGG}$. Choosing the IGG as $\mathbb{Z}_2$ one obtains
\begin{align}
&G_{T_y}(\mathbf{r}) G_{T_z}(\mathbf{r}-\hat{e}_y)G_{T_y}^{-1}(\mathbf{r}-\hat{e}_z)G_{T_z}^{-1}(\mathbf{r}) = \pm \tau^0 = \eta_{z_y}\tau^0\notag\\
&\hspace*{-5pt}\implies G_{T_y}(\mathbf{r})=\eta_{z_y}G_{T_y}(\mathbf{r}-\hat{e}_z)\;,
\end{align}
where $\eta_{z_y}=\pm 1$ and it was used that $G_{T_z}(\mathbf{r})=\tau^0$ for all $\mathbf{r}$. This equation is solved by $G_{T_y}(\mathbf{r})=\eta_{z_y}^{z}g_{T_y}$ where $g_{T_y}$ is a site-independent $SU(2)$ matrix. On the other hand, we know that $G_{T_y}(\mathbf{r}_{xy})=\tau^0$ and thus $G_{T_y}(\mathbf{r})=\eta_{z_y}^{z-{z_0}}\tau^0$. In complete analogy one finds the projective representation for translations in $x$ direction $G_{T_x}(\mathbf{r})=\eta_{z_x}^{z-z_0}\eta_{y_x}^{y-y_0}\tau^0$. Since $\mathbf{r}_0$ is arbitrary we can choose it as the origin $\mathbf{r}_0 = (0, 0, 0)$. 

\section{Projective implementation of permutation $P$}\label{ap:gauge_example}
To demonstrate how the defining equations for the PSGs on the sc lattice listed in Eq.~(\ref{eq:algebra}) are obtained, we determine, as an example, the projective action of $P$ and its consequences for the implementation of translations $T_\mu$. In the gauge derived in Appendix~\ref{ap:gauge} the projective implementation of the point group elements can be determined by inspection of group actions which map onto the identity. Since the representations of the translations are already fixed it is convenient to start with the mutual relations between point group elements and translations. For permutation $P$ this yields
\begin{align*}
G_{P} P G_{T_x} T_x P^{-1}G_{P}^{-1} T_y^{-1} G_{T_y}^{-1} &\in \text{IGG}\;.
\end{align*}
Similar expressions can be obtained under cyclic permutation $x\rightarrow y\rightarrow z\rightarrow x$. For an IGG given by $\mathbb{Z}_2$ one obtains
\begin{align*}
& G_{P}(x,y,z)G_{T_x}(y,z,x)G_{P}^{-1}(x,y-1,z)G_{T_y}^{-1}(x,y,z)\\
&\hspace*{-2.5pt}=\eta_{y_P}\tau^0\hspace*{-1pt}\implies\hspace*{-1pt}G_{P}(x,y,z) = \eta^{z}_{y_x}\eta^{x}_{z_x}\eta^{z}_{z_y}\eta_{y_P}G_{P}(x,y-1,z)\;,
\end{align*}
and similarly
\begin{align*}
& G_{P}(x,y,z) = \eta^{z}_{z_x}\eta^{y}_{y_x}\eta_{x_P}G_{P}(x-1,y,z)\;, \\
& G_{P}(x,y,z) = \eta^{x}_{z_y}\eta_{z_P}G_{P}(x,y,z-1)\;.
\end{align*}
To find a solution to these equations one constructs relations between $G_{P}({\mathbf r})$ along elementary loops including the origin using the known action of the translations. These loop operations serve as consistency conditions as they are equal to an identity operation. As an example, we consider $G_{P}(x,y,z)$ along a loop in the $x$-$y$ plane:
\begin{align*}
& G_P(0,0,0)=\eta_{x_P}G_P(1,0,0)\;,  \\
& G_P(1,0,0)=\eta_{z_x} \eta_{y_P} G_P(1,1,0) = \eta_{x_P} G_P(0,0,0)\;,\\
& G_P(1,1,0)=\eta_{y_x}\eta_{x_P} G_P(0,1,0) = \eta_{z_x} \eta_{y_P} \eta_{x_P} G_P(0,0,0)\;,\\
& G_P(0,1,0)=\eta_{y_P} G_P(0,0,0) = \eta_{y_x} \eta_{z_x} \eta_{y_P} G_P(0,0,0)\;.
\end{align*}
The last equation shows that the symmetry $P$ requires $\eta_{y_x} = \eta_{z_x}$. Repeating this process in the other planes reveals that there is only one sign parameter for translations, $\eta_{y_x} = \eta_{z_x} = \eta_{z_y}\equiv \eta_X$. Relations of this type also allow one to determine the spatial dependence of $G_{P}({\mathbf r})$. Fixing the projective representation at the origin, $G_P(0,0,0)\equiv g_P$, yields the unique solution
\begin{align*}
G_P(\mathbf{r})=\eta_X^{x(y+z)}\eta^x_{x_P}\eta^y_{y_P}\eta^z_{z_P}g_P\;.
\end{align*}
The projective representations of the other point group generators can be similarly decomposed into site-dependent sign factors $\eta$ and site-independent $SU(2)$ matrices $g$. These matrices are further specified by exploiting the mutual relations between different point group generators. This leads to the full set of algebraic conditions listed in Eq.~\eqref{eq:algebra}. The corresponding gauge-inequivalent solutions are presented in Appendix~\ref{ap:irreps}.

\section{Gauge-inequivalent PSG representations for the sc, bcc, and fcc lattices}\label{ap:irreps}
In Table~\ref{tab:reps} we list all sets of gauge-inequivalent representation matrices $g_{\mathcal{O}}$ for the point group generators $\mathcal{O}$ of the sc lattice. The matrices corresponding to $\Pi_z$ and $\Pi_y$ can only be represented trivially, $g_{\Pi_{z}}=g_{\Pi_{y}}=\tau^0$. There are 21 different solutions for the remaining matrices $g_\mathcal{T}$, $g_P$, $g_I$, $g_{\Pi_{xy}}$. For each solution the sign factors $\eta_\mathcal{O}=\pm 1$ complete the PSG representation. Note, however, that the case $g_\mathcal{T}=\tau^0$ and $\eta_\mathcal{T}=1$ does not lead to finite mean-field ans\"atze. For the fcc lattice the additional translations $t_1$ and $t_2$ can only have a trivial matrix structure, $g_{t_1}=g_{t_2}=\tau^0$. The representation matrices are, therefore, the same as for the sc lattice (see Table~\ref{tab:reps}). For the bcc lattice all 59 gauge inequivalent solutions are shown in Table~\ref{tab:reps_bcc} where, in addition to $g_\mathcal{T}$, $g_P$, $g_I$, $g_{\Pi_{xy}}$ the possible solutions for $g_t$ are specified.
\begin{center}
\setlength{\tabcolsep}{0.2cm}
\begin{table*}[h]
\begin{minipage}{0.3\linewidth}
\begin{tabular}{ >{$}c<{$}  >{$}c<{$}  >{$}c<{$}  >{$}c<{$}  >{$}c<{$} }
\hline\hline
\text{PSG} & g_\mathcal{T} & g_P & g_I & g_{\Pi_{xy}} \\
\hline
1 & \tau^0 & \tau^0 & \tau^0 & \tau^0 \\
2 & \tau^0 & \tau^0 & i\tau^2 & \tau^0 \\
3 & \tau^0 & \tau^0 & \tau^0 & i\tau^2 \\
4 & \tau^0 & \tau^0 & i\tau^2 & i\tau^2 \\
5 & \tau^0 & \tau^0 & i\tau^2 & i\tau^3 \\
6 & \tau^0 & e^{i\frac{\pi}{3}\tau^2} & \tau^0 & i\tau^3 \\
7 & \tau^0 & e^{i\frac{\pi}{3}\tau^2} & i\tau^2 & i\tau^3 \\
\hline\hline
\end{tabular}
\end{minipage}
\begin{minipage}{0.3\linewidth}
\begin{tabular}{ >{$}c<{$}  >{$}c<{$}  >{$}c<{$}  >{$}c<{$}  >{$}c<{$} }
\hline\hline
\text{PSG} & g_\mathcal{T} & g_P & g_I & g_{\Pi_{xy}} \\
\hline
8 & \tau^0 & e^{i\frac{2\pi}{3}\tau^2} & \tau^0 & i\tau^3 \\
9 & \tau^0 & e^{i\frac{2\pi}{3}\tau^2} & i\tau^2 & i\tau^3 \\
10 & i\tau^2 & \tau^0 & \tau^0 & \tau^0 \\
11 & i\tau^2 & \tau^0 & i\tau^2 & \tau^0 \\
12 & i\tau^2 & \tau^0 & i\tau^3 & \tau^0 \\
13 & i\tau^2 & \tau^0 & \tau^0 & i\tau^2 \\
14 & i\tau^2 & \tau^0 & \tau^0 & i\tau^3\\
\hline\hline
\end{tabular}
\end{minipage}
\begin{minipage}{0.3\linewidth}
\begin{tabular}{ >{$}c<{$}  >{$}c<{$}  >{$}c<{$}  >{$}c<{$}  >{$}c<{$} }
\hline\hline
\text{PSG} & g_\mathcal{T} & g_P & g_I & g_{\Pi_{xy}} \\
\hline
15 & i\tau^2 & \tau^0 & i\tau^2 & i\tau^2 \\
16 & i\tau^2 & \tau^0 & i\tau^3 & i\tau^2 \\
17 & i\tau^2 & \tau^0 & i\tau^3 & i\tau^3 \\
18 & i\tau^2 & e^{i\frac{\pi}{3}\tau^2} & \tau^0 & i\tau^3 \\
19 & i\tau^2 & e^{i\frac{\pi}{3}\tau^2} & i\tau^2 & i\tau^3 \\
20 & i\tau^2 & e^{i\frac{2\pi}{3}\tau^2} & \tau^0 & i\tau^3 \\
21 & i\tau^2 & e^{i\frac{2\pi}{3}\tau^2} & i\tau^2 & i\tau^3 \\
\hline\hline
\end{tabular}
\end{minipage}

\caption{Projective representation matrices $g_\mathcal{T}$, $g_P$, $g_I$, $g_{\Pi_{xy}}$ for the sc and fcc lattices.}\label{tab:reps}
\end{table*}
\end{center}

\begin{center}
\begin{table*}[h]
\setlength{\tabcolsep}{0.1cm}
\begin{minipage}{0.3\linewidth}
\begin{tabular}{ >{$}c<{$}  >{$}c<{$}  >{$}c<{$}  >{$}c<{$}  >{$}c<{$} >{$}c<{$} }
\hline\hline
\text{PSG} & g_\mathcal{T} & g_P & g_I & g_{\Pi_{xy}} & g_{t} \\
\hline
1 & \tau^0 / i\tau^2  & \tau^0 & \tau^0 & \tau^0 & \tau^0 \\
2 & \tau^0 / i\tau^2 & \tau^0 & i\tau^2 & \tau^0 & \tau^0 \\
3 & \tau^0 / i\tau^2 & \tau^0 & \tau^0 & i\tau^2 & \tau^0 \\
4 & \tau^0 / i\tau^2 & \tau^0 & \tau^0 & \tau^0 & i\tau^2 \\
5 & \tau^0 / i\tau^2 & \tau^0 & \tau^0 & i\tau^2 & i\tau^2 \\
6 & \tau^0 / i\tau^2 & \tau^0 & i\tau^2 & i\tau^2 & \tau^0 \\
7 & \tau^0 / i\tau^2 & \tau^0 & i\tau^2 & \tau^0 & i\tau^2 \\
8 & \tau^0 / i\tau^2 & \tau^0 & i\tau^2 & i\tau^2 & i\tau^2 \\
9 & \tau^0 / i\tau^2 & \tau^0 & i\tau^2 & i\tau^3 & \tau^0 \\
10 & \tau^0 / i\tau^2 & \tau^0 & i\tau^2 & \tau^0 & i\tau^3 \\
11 & \tau^0 / i\tau^2 & \tau^0 & \tau^0 & i\tau^2 & i\tau^3 \\
12 & \tau^0 / i\tau^2 & \tau^0 & i\tau^2 & i\tau^3 & i\tau^3 \\
\hline\hline
\end{tabular}
\end{minipage}\quad
\begin{minipage}{0.3\linewidth}
\begin{tabular}{ >{$}c<{$}  >{$}c<{$}  >{$}c<{$}  >{$}c<{$}  >{$}c<{$} >{$}c<{$} }
\hline\hline
\text{PSG} & g_\mathcal{T} & g_P & g_I & g_{\Pi_{xy}} & g_{t} \\
\hline
13 & \tau^0 / i\tau^2 & \tau^0 & i\tau^2 & i\tau^2 & i\tau^3 \\
14 & \tau^0 / i\tau^2 & \tau^0 & i\tau^2 & i\tau^3 & i\tau^2 \\
15 & \tau^0 / i\tau^2 & \tau^0 & i\tau^2 & i\tau^3 & i\tau^1 \\
16 & i\tau^2 & \tau^0 & i\tau^3 & \tau^0 & \tau^0 \\
17 & i\tau^2 & \tau^0 & \tau^0 & i\tau^3 & \tau^0 \\
18 & i\tau^2 & \tau^0 & \tau^0 & \tau^0 & i\tau^3 \\
19 & i\tau^2 & \tau^0 & \tau^0 & i\tau^3 & i\tau^3 \\
20 & i\tau^2 & \tau^0 & \tau^0 & i\tau^3 & i\tau^1 \\
21 & i\tau^2 & \tau^0 & i\tau^3 & i\tau^2 & \tau^0 \\
22 & i\tau^2 & \tau^0 & i\tau^3 & \tau^0 & i\tau^2 \\
23 & i\tau^2 & \tau^0 & i\tau^3 & i\tau^2 & i\tau^2 \\
24 & i\tau^2 & \tau^0 & i\tau^3 & i\tau^3 & \tau^0 \\
\hline\hline
\end{tabular}
\end{minipage}\quad
\begin{minipage}{0.3\linewidth}
\begin{tabular}{ >{$}c<{$}  >{$}c<{$}  >{$}c<{$}  >{$}c<{$}  >{$}c<{$} >{$}c<{$} }
\hline\hline
\text{PSG} & g_\mathcal{T} & g_P & g_I & g_{\Pi_{xy}} & g_{t} \\
\hline
25 & i\tau^2 & \tau^0 & i\tau^3 & \tau^0 & i\tau^3 \\
26 & i\tau^2 & \tau^0 & i\tau^3 & i\tau^3 & i\tau^3 \\
27 & i\tau^2 & \tau^0 & i\tau^3 & i\tau^3 & i\tau^1 \\
28 & i\tau^2 & \tau^0 & i\tau^3 & i\tau^1 & i\tau^1 \\
29 & \tau^0 / i\tau^2 & e^{i\frac{\pi}{3}\tau^2} & \tau^0 & i\tau^3 & \tau^0 \\
30 & \tau^0 / i\tau^2 & e^{i\frac{\pi}{3}\tau^2} & \tau^0 & i\tau^3 & i\tau^2 \\
31 & \tau^0 / i\tau^2 & e^{i\frac{\pi}{3}\tau^2} & i\tau^2 & i\tau^3 & \tau^0 \\
32 & \tau^0 / i\tau^2 & e^{i\frac{\pi}{3}\tau^2} & i\tau^2 & i\tau^3 & i\tau^2 \\
33 & \tau^0 / i\tau^2 & e^{i\frac{2\pi}{3}\tau^2} & \tau^0 & i\tau^3 & \tau^0 \\
34 & \tau^0 / i\tau^2 & e^{i\frac{2\pi}{3}\tau^2} & \tau^0 & i\tau^3 & i\tau^2 \\
35 & \tau^0 / i\tau^2 & e^{i\frac{2\pi}{3}\tau^2} & i\tau^2 & i\tau^3 & \tau^0 \\
36 & \tau^0 / i\tau^2 & e^{i\frac{2\pi}{3}\tau^2} & i\tau^2 & i\tau^3 & i\tau^2 \\
\hline\hline
\end{tabular}
\end{minipage}
\caption{Projective representation matrices $g_\mathcal{T}$, $g_P$, $g_I$, $g_{\Pi_{xy}}$, $g_t$ for the bcc lattice. The notation $\tau^0/ i\tau^2$ indicates that $g_{\mathcal{T}}$ can either be represented by $\tau^0$ or $i\tau^2$.}\label{tab:reps_bcc}
\end{table*}
\end{center}

\section{Algebraic PSGs of the bcc lattice}\label{ap:bcc}
Here, we present further details about our procedure to determine the algebraic PSGs for the bcc lattice. The fcc case may be treated similarly. As explained in the main text, we use two distinct sc lattices and merge them into a bcc lattice. The two cubic sublattices are denoted $A=\left\lbrace (x,y,z)| x,y,z \in \mathbb{Z} \right\rbrace$ and $B=\left\lbrace(x+1/2,y+1/2,z+1/2)|x,y,z \in \mathbb{Z} \right\rbrace$. On each sublattice we have a complete description of the symmetry representations given by Eq.~\eqref{eq:algebra}. To distinguish between the two sublattices we add an extra label in the projective representations $G_{\mathcal{S}}^A(\mathbf{r}\in A)$ and $G_{\mathcal{S}}^B(\mathbf{r}\in B)$. The implementations of symmetries on sublattice $A$ are done in complete analogy to the sc lattice while on sublattice $B$ one needs to define a reference site $\mathbf{r}^B_0 = (1/2, 1/2, 1/2)$ which remains invariant under point group operations. The symmetry operation $t$ connects both sublattices.

To determine the projective action of $t$ we consider the operation $T^{-1}_xtT_x t^{-1} = id$ which moves a given site ${\mathbf r}$ along a closed path. Including the associated gauge transformations this relation reads $T^{-1}_x(G_{T_x})^{-1}G_{t}t (G_{T_x})T_x t^{-1}(G_{t})^{-1}\in \text{IGG}$ which results in a condition for the projective representation of $t$ on sublattice $A$:
\begin{align*}
&(G^A_{T_x})^{-1}(x+1,y,z)G^A_{t}(x+1, y, z)\times \\
&\hspace*{-1.5pt}\times G^B_{T_x}(x+1/2, y-1/2, z-1/2) (G^A_{t})^{-1}(x, y, z)= \eta^{A}_{t_x}\tau^0 \\
&\hspace*{-4pt}\implies G^A_{t}(x, y, z)= (\eta^A_{X}\eta^B_{X})^{y+z}\eta^{A}_{t_x}G^A_{t}(x+1, y, z)\;.
\end{align*}
Note that with the above definition of sublattices $x,y,z \in \mathbb{Z}$ the exponents of the $\eta$ parameters always take integer values. Similarly, one finds conditions involving translations $T_\mu$ along the other cartesian directions,
\begin{align*}
& G^A_{t}(x, y, z)= (\eta^A_{X}\eta^B_{X})^{z}\eta^{A}_{t_y}G^A_{t}(x, y+1, z)\;, \\
& G^A_{t}(x, y, z)= \eta^{A}_{t_z}G^A_{t}(x, y, z+1)\;.
\end{align*}
Equivalently, on sublattice $B$ one finds
\begin{align*}
&G^B_{t}(x+1/2, y+1/2, z+1/2)= \\
&(\eta^A_{X}\eta^B_{X})^{y+z}\eta^{B}_{t_x}G^B_{t}(x+3/2, y+1/2, z+1/2)\;, \\
&G^B_{t}(x+1/2, y+1/2, z+1/2)= \\
&(\eta^A_{X}\eta^B_{X})^{z}\eta^{B}_{t_y}G^B_{t}(x+1/2, y+3/2, z+1/2)\;, \\
&G^B_{t}(x+1/2, y+1/2, z+1/2)= \\
&\eta^{B}_{t_z}G^B_{t}(x+1/2, y+1/2, z+3/2)\;.
\end{align*}
Following the line of arguments of Appendix~\ref{ap:gauge_example}, closed loops of symmetry operations provide consistency conditions which reveal that a solution can only exist for $\eta^A_{X}=\eta^B_{X}\equiv \eta_{X}$. It further follows
\begin{align*}
&G^A_t(\mathbf{r})=(\eta^{A}_{t_x})^x(\eta^{A}_{t_y})^y(\eta^{A}_{t_z})^zg^A_t\;, \notag \\
&G^B_t(\mathbf{r})=(\eta^{B}_{t_x})^x(\eta^{B}_{t_y})^y(\eta^{B}_{t_z})^zg^B_t\;.
\end{align*} 
Relations between the two sublattices can be found using $t^2 = T_zT_yT_x$ which yields $\eta^A_{t_x} = \eta^B_{t_x} \equiv \eta_{t_x}$, $\eta^A_{t_y} = \eta_X \eta^B_{t_y} \equiv \eta_{t_y}$ and $\eta^A_{t_z} = \eta^B_{t_z} \equiv \eta_{t_z}$. Furthermore, the site-independent matrices $g^A_t$, $g^B_t$ need to fulfill $g^A_t g^B_t=g^B_t g^A_t=\pm \tau^0$ such that we can define $g^A_t = \pm g^B_t \equiv g_t$ with $g_t^2=\pm \tau^0$.

In the next step we include lattice inversion $I$. We again note that in the initial implementation of point group symmetries, inversion on sublattice $B$, referred to as $I^B$, leaves the reference site $\mathbf{r}^B_0 = (1/2, 1/2, 1/2)$ invariant:
\begin{align*}
&I^B(x+1/2, y+1/2, z+1/2)\notag \\
&{=}\;(-x+1/2, -y+1/2, -z+1/2)\;.
\end{align*}
It is still convenient to define an inversion $I$ for the entire bcc lattice in the usual way where one site ${\mathbf r}_0=(0,0,0)$ is globally left invariant. This can be achieved via the relation between $I$ and $I^B$ on sublattice $B$,
\begin{align*}
I^B({\mathbf r}\in B)=T_xT_yT_zI({\mathbf r}\in B)\;,
\end{align*}
which implies
\begin{align*}
G_{I^B}I^B({\mathbf r}\in B)=G_{T_x}T_x G_{T_y} T_y G_{T_z} T_z G_{I} I({\mathbf r}\in B)\;.
\end{align*}
(Note that similar distinctions between the action on sublattice $B$ and the global action also have to be made for the generators $\Pi_z, \Pi_y$ and $\Pi_{xy}$.) Exploiting the algebraic relation $I^{-1}t^{-1}I^Bt({\mathbf r}\in A)=I^{-1}t^{-1}T_x T_y T_z It({\mathbf r}\in A)=id$ between inversion $I$ and translations $T_\mu$, $t$ leads to $\eta^A_I=\eta^B_I$ and $\eta^A_{t_y} \eta^B_{t_y} \eta^A_I \eta^B_I =1$. In combination with the previous result $\eta^A_{t_y} = \eta_X \eta^B_{t_y}$ one obtains the important finding $\eta_X=1$. This means that all gauge transformations associated with translations $T_\mu$ are now trivially represented by $\tau^0$ such that all $G_{T_\mu}$ in the relations between $G_{I^B}$ and $G_{I}$ drop out. Furthermore, the conditions $Pt=tP$ and $\Pi_{xy}T_zt^{-1}\Pi_{xy}t=id$ connect the sign factors corresponding to different directions $\eta_{t_x}=\eta_{t_y}=\eta_{t_z}\equiv \eta_t$.

Having derived the sign structure of the gauge transformations associated with translations, we now turn to the matrix structure. Exploiting the fact that translations $T_\mu$ have a trivial projective implementation one finds 
\begin{align}
&g_t^{-1}g^A_I g_t = \pm g^{B}_I\;, \notag \\
&g_t^{-1}g^A_{\mathcal{T}} g_t = \pm g^{B}_{\mathcal{T}}\;, \notag \\
&g_t^{-1}g^A_{\Pi_{xy}} g_t = \pm g^{B}_{\Pi_{xy}}\;, \notag \\
&g_t^{-1}g^A_P g_t = \pm g^{B}_P\;,\label{gagb}
\end{align}
where, initially, one would assume that each of the two sets $g^A_\mathcal{S}$ and $g^B_\mathcal{S}$ can be independently given by one line of Table~\ref{tab:reps}. It is, however, easy to see that the representations need to be identical, $g^A_\mathcal{S}=\pm g^B_\mathcal{S}$ on the two sublattices (up to an irrelevant sign). Otherwise, Eq.~(\ref{gagb}) would imply that $g_t$ transforms between two different PSGs on the sc lattice. Since, by construction, different PSGs are gauge-inequivalent, this is not possible. Thus, we conclude that the classification of PSGs for the sc lattice can be reused for both sublattices of the bcc lattice where one finds $\eta_X=1$ and an additional generator $G_t(\mathbf{r})=\eta_{t}^{x+y+z}g_t$ needs to be considered.

\section{Dynamical spin structure factor}\label{ap:structure_factor}
The dynamical spin structure factor investigated in the main text,
\begin{align}\label{eq:stru_general}
\mathcal{S}^{\mu\nu}(\mathbf{q}, \omega) = \int^{\infty}_{-\infty} \frac{dt}{2\pi} e^{i\omega t} \frac{1}{N}\sum_{\mathbf{rr'}}e^{i\mathbf{q}(\mathbf{r}-\mathbf{r'})}\left\langle S^\mu_\mathbf{r}(t) S^\nu_\mathbf{r'}(0) \right\rangle\;,
\end{align}
is a measure of the system's magnetic excitation spectrum as a function of momentum $\mathbf{q}$ and frequency $\omega$ and is directly accessible via inelastic neutron scattering. Since in our systems we always assume spin-rotation invariance it suffices to consider the longitudinal components $\mu=\nu=z$ only. In the fermionic representation applied here, the dynamical spin structure factor can be expressed as
\begin{align}\label{eq:stru_ferm}
&\mathcal{S}^{zz}({\mathbf q,\omega})=\frac{\pi}{4}\sum_{a,b}\int_{\text{BZ}}\frac{d^3 k}{(2\pi)^3}f({\mathbf k},{\mathbf q},a,b)\notag\\
&\times[n_a({\mathbf k})-n_b({\mathbf k}+{\mathbf q})]\delta(\epsilon_b({\mathbf k}+{\mathbf q})-\epsilon_a({\mathbf k})-\omega)\;.
\end{align}
Here, $\epsilon_a$ is an eigenenergy of Eq.~\eqref{eq:fermion_hamiltonian} with $n_a$ the occupation number of the energy band labeled by an index $a$ and the function $f({\mathbf k},{\mathbf q},a,b)$ describes the overlap between different eigenstates $\psi_a(\mathbf{k})$ defined by
\begin{align}
f({\mathbf k},{\mathbf q},a,b)= \left|\psi^*_a(\mathbf{k})\psi_b(\mathbf{k+\mathbf{q}}) \right|^2\;.
\end{align}

\section{Compendium of short-ranged mean-field ans\"atze}
In the following Tables~ \ref{tab:sc_summary}, \ref{tab:bcc_summary}, and \ref{tab:fcc_summary} we list all possible short-range mean-field ans\"atze (including mean-field terms up to third neighbors) for the sc, bcc and fcc lattices and also provide the projective implementations of symmetries.

\begin{table*}[h]
\setlength{\tabcolsep}{0.1cm}
\begin{tabular}{ >{$}c<{$}  >{$}c<{$}  >{$}c<{$}  >{$}c<{$}  >{$}c<{$}  >{$}c<{$}  >{$}c<{$}  >{$}c<{$}  >{$}c<{$}  >{$}c<{$}  >{$}c<{$} }
\hline\hline
& \eta_\mathcal{T} g_\mathcal{T} & \eta_{P} g_P & \eta_I g_I & \eta_{X} g_X & \eta_{\Pi_{xy}} g_{\Pi_{xy}} & u_{\delta \mathbf{r}_1} & u_{\delta \mathbf{r}_2} & u_{\delta \mathbf{r}_3} & a_\mu &IGG\\
\hline
& +i\tau^2 & +\tau^0 & +\tau^0 & +\tau^0 & +\tau^0 & \chi_1 \tau^3 & \textcolor{red}{\chi_2 \tau^3 + \Delta_2 \tau^1} & \textcolor{blue}{\chi_3 \tau^3 + \Delta_3 \tau^1} & a_3 (a_1) & SU(2)/\textcolor{red}{U(1)}/\textcolor{blue}{\mathbb{Z}_2}\\
& +i\tau^2 & +\tau^0 & +\tau^0/ +i\tau^3 & +\tau^0 & +\tau^0/ +i\tau^3 & \chi_1 \tau^3 & \textcolor{red}{\chi_2 \tau^3} & \chi_3 \tau^3 & a_3 & SU(2)/\textcolor{red}{U(1)} \\
& -i\tau^3 & +\tau^0 & +\tau^0/ -i\tau^2 & +\tau^0 & +\tau^0 & \chi_1 \tau^3 & \textcolor{red}{\Delta_2 \tau^2} & \chi_3 \tau^3 & a_2 & SU(2)/\textcolor{red}{U(1)} \\
& -i\tau^3 & -\tau^0 & +\tau^0/ -i\tau^2 & +\tau^0 & -i\tau^2 & \chi_1 \tau^3 & \textcolor{red}{\Delta_2 \tau^2} & \chi_3 \tau^3 & a_2 & SU(2)/\textcolor{red}{U(1)} \\
\text{SC 1} & -\tau^0 & +\tau^0 & +\tau^0 & +\tau^0 & +\tau^0 & \chi_1 \tau^3 &-& \textcolor{red}{\chi_3 \tau^3 + \Delta_3 \tau^1+\Delta_{3}^{'}\tau^{2}} &-&SU(2)/\textcolor{red}{U(1)}\\
& -\tau^0 & +\tau^0 & -i\tau^2 & +\tau^0 & +\tau^0 & \chi_1 \tau^3 &-& \textcolor{red}{\chi_3 \tau^3 + \Delta_3 \tau^1} &-&SU(2)/\textcolor{red}{U(1)}\\
& -\tau^0/ +i\tau^2 & -\tau^0 & +\tau^0/ -i\tau^2 & +\tau^0 & -i\tau^2 & \chi_1 \tau^3 &-& \textcolor{red}{\chi_3 \tau^3 + \Delta_3 \tau^1} &-&SU(2)/\textcolor{red}{U(1)}\\
& +i\tau^2 & +\tau^0 & -i\tau^2 & +\tau^0 & +\tau^0 & \chi_1 \tau^3 &-& \textcolor{red}{\chi_3 \tau^3 + \Delta_3 \tau^1} &-&SU(2)/\textcolor{red}{U(1)}\\
& -\tau^0 & +\tau^0 & -i\tau^2 & +\tau^0 & +i\tau^3 & \chi_1 \tau^3 &-& \chi_3 \tau^3 &-&SU(2)\\
& +i\tau^2 & -\tau^0 & +i\tau^3 & +\tau^0 & -i\tau^2 & \chi_1 \tau^3 &-& \chi_3 \tau^3 &-&SU(2)\\
\hline
& +i\tau^2 & -\tau^0 & +\tau^0 & -\tau^0 & -i\tau^2 & \chi_1 \tau^3 & \textcolor{red}{\chi_2 \tau^3 + \Delta_2 \tau^1} &-& - & SU(2)/\textcolor{red}{U(1)} \\
\text{SC 2}& +i\tau^2 & -\tau^0 & +i\tau^3 & -\tau^0 & -i\tau^2 & \chi_1 \tau^3 & \textcolor{red}{\chi_2 \tau^3} &-& - & SU(2)/\textcolor{red}{U(1)} \\
& -i\tau^3 & +\tau^0 & +\tau^0/ -i\tau^2 & -\tau^0 & +i\tau^3 & \chi_1 \tau^3 & \textcolor{red}{\Delta_2 \tau^2} &-&-&SU(2)/\textcolor{red}{U(1)} \\
\hline\hline
\end{tabular}
\caption{Possible short-range PSG representations on the sc lattice and their corresponding mean-field ans\"atze. The color code (red/blue) indicates which term is responsible for breaking the IGG down to $U(1)/\mathds{Z}_2$. Note that in the second line at least one of the matrices $g_I$,$g_{\Pi_{xy}}$ must be given by $i\tau^3$.}\label{tab:sc_summary}
\end{table*}

\begin{table*}[h]
\setlength{\tabcolsep}{0.14cm}
\begin{tabular}{ >{$}c<{$}  >{$}c<{$}  >{$}c<{$}  >{$}c<{$}  >{$}c<{$}   >{$}c<{$}  >{$}c<{$}   >{$}c<{$}  >{$}c<{$}  >{$}c<{$}  >{$}c<{$} }
\hline\hline
 & \eta_{\mathcal{T}}g_\mathcal{T} & \eta_P g_P & \eta_I g_I & \eta_{\Pi_{xy}} g_{\Pi_{xy}} & \eta_t g_{t} & u_{\delta \mathbf{r}_1} & u_{\delta \mathbf{r}_2} & u_{\delta \mathbf{r}_3} & a_\mu & IGG \\
\hline
\text{BCC 1}& +i\tau^2  & +\tau^0 & +\tau^0 & +\tau^0 & +\tau^0 & \chi_1\tau^3 & \textcolor{red}{\chi_2\tau^3 + \Delta_2 \tau^1} & \textcolor{blue}{\chi_3\tau^3 + \Delta_3 \tau^1} & a_3 (a_1) & SU(2)/\textcolor{red}{U(1)}/\textcolor{blue}{\mathbb{Z}_2} \\
& +i\tau^2  & +\tau^0 & +\tau^0/+i\tau^3 & +\tau^0/+i\tau^3 & +\tau^0/+i\tau^3 & \chi_1\tau^3 & \textcolor{red}{\chi_2\tau^3} & \chi_3\tau^3 & a_3 & SU(2)/\textcolor{red}{U(1)} \\
\hline
\text{BCC 2}& +i\tau^2  & -\tau^0 & +\tau^0 & +\tau^0 & +\tau^0 & \chi_1\tau^3 & - & \textcolor{blue}{\chi_3\tau^3 + \Delta_3 \tau^1} & a_3 (a_1) & SU(2)/\textcolor{blue}{\mathbb{Z}_2} \\
& +i\tau^2  & -\tau^0 & +\tau^0/+i\tau^3 & +\tau^0/+i\tau^3 & +\tau^0/+i\tau^3 & \chi_1\tau^3 & - & \textcolor{red}{\chi_3\tau^3} & a_3 & SU(2)/\textcolor{red}{U(1)} \\
\hline\hline
\end{tabular}

\caption{Possible short-range PSG representations on the bcc lattice and their corresponding mean-field ans\"atze. The color code (red/blue) indicates which term is responsible for breaking the IGG down to $U(1)/\mathds{Z}_2$. Note that in the lines with entries $+\tau^0/+i\tau^3$ both $+\tau^0$ and $+i\tau^3$ are possible, but at least one of these matrices must be given by $+i\tau^3$.}\label{tab:bcc_summary}
\end{table*}

\begin{table*}[h]
\setlength{\tabcolsep}{0.17cm}
\begin{tabular}{ >{$}c<{$}  >{$}c<{$}  >{$}c<{$}  >{$}c<{$}  >{$}c<{$}   >{$}c<{$}  >{$}c<{$}   >{$}c<{$}  >{$}c<{$}  >{$}c<{$} }
\hline\hline
 & \eta_{\mathcal{T}}g_\mathcal{T} & \eta_P g_P & \eta_I g_I & \eta_{\Pi_{xy}} g_{\Pi_{xy}} & u_{\delta \mathbf{r}_1} & u_{\delta \mathbf{r}_2} & u_{\delta \mathbf{r}_3} & a_\mu & IGG \\
\hline
\text{FCC 1}& +i\tau^2  & +\tau^0 & +\tau^0 & +\tau^0 & \chi_1\tau^3 & \textcolor{red}{\chi_2\tau^3 + \Delta_2 \tau^1} & \chi_3\tau^3 + \Delta_3 \tau^1 & a_3 (a_1) & U(1)/\textcolor{red}{\mathbb{Z}_2} \\
& +i\tau^2  & +\tau^0 & +\tau^0/+i\tau^3 & +\tau^0/+i\tau^3 & \chi_1\tau^3 & \chi_2\tau^3 & \chi_3\tau^3 & a_3 & U(1) \\
\hline
\text{FCC 2} & +i\tau^2  & +e^{i\frac{\pi}{3}\tau^2} & +\tau^0 & +i\tau^3 & \textcolor{blue}{\chi_1 f(\delta \mathbf{r}_1, \tau^1, \tau^3)} & - & \chi_3 f(\delta \mathbf{r}_3, \tau^1, \tau^3) & - & \mathbb{Z}_2 \\
& +i\tau^2  & +e^{i\frac{2\pi}{3}\tau^2} & +\tau^0 & +i\tau^3 & \textcolor{blue}{\chi_1 g(\delta \mathbf{r}_1, \tau^1, \tau^3)} & - & \chi_3 g(\delta \mathbf{r}_3, \tau^1, \tau^3) & - & \mathbb{Z}_2 \\
\hline\hline
\end{tabular}
\caption{Possible short-range PSG representations on the fcc lattice and their corresponding mean-field ans\"atze. The color code (red/blue) indicates which term is responsible for breaking the IGG down to $U(1)/\mathds{Z}_2$. The mean-field Hamiltonian in the FCC 2 case has a non-trivial matrix structure denoted by $f(\delta \mathbf{r}_1, \tau^1, \tau^3)$ for nearest neighbor amplitudes [see Eq.~(\ref{eq:fcc_2_1})] and $f(\delta \mathbf{r}_3, \tau^1, \tau^3)$ for third neighbor amplitudes [see Eq.~(\ref{eq:fcc_2_3})]. The functions $g$ are similar but the axes are permuted according to $(x,y,z)\rightarrow P(x,y,z)$. Note that in the line with entries $+\tau^0/+i\tau^3$ both $+\tau^0$ and $+i\tau^3$ are possible, but at least one of these matrices must be given by $+i\tau^3$.}\label{tab:fcc_summary}
\end{table*}

%\end{appendices}
%\bibliographystyle{revtex}
\bibliography{refs}
\end{document}